\documentclass[12pt,letterpaper]{article}

\usepackage[margin=1in]{geometry}
\usepackage{graphicx}
\usepackage{caption}
\usepackage{float}
\usepackage{booktabs}
\usepackage{array}
\usepackage{multirow}
\usepackage{adjustbox}
\usepackage{threeparttable}
\usepackage[flushleft]{threeparttablex}
\usepackage{amsmath,amssymb}
\usepackage[breaklinks]{hyperref}
\usepackage{natbib}
\usepackage{setspace}

\hypersetup{
  colorlinks=true,
  linkcolor=blue,
  citecolor=blue,
  urlcolor=blue,
}

\bibpunct{(}{)}{;}{a}{,}{,}

\newenvironment{tableresize}{%
  \small
  \begin{adjustbox}{max width=\textwidth}%
}{%
  \end{adjustbox}%
}

\begin{document}
\thispagestyle{empty}
\singlespacing

\vspace*{0.5in}

\begin{center}
\begin{tabular}{c}
  {\Large\bfseries Advertising Spillovers in Mobile Apps:} \\[0.35em]
  {\Large\bfseries Evidence from Ad Shutoffs and Store Rankings}
\end{tabular}

\vspace{1.25em}

\begin{tabular}[t]{@{}c@{\hspace{1.5em}}c@{\hspace{1.5em}}c@{}}
  Harang Ju\footnotemark[1]
  & Michael Zhao\footnotemark[1]
  & Sinan Aral\footnotemark[2] \\[0.4em]
  {\small Johns Hopkins Carey}
  & {\small DoorDash}
  & {\small MIT Sloan School of} \\
  {\small Business School}
  & {\small\phantom{x}}
  & {\small Management} \\[0.25em]
  {\small\texttt{harang@jhu.edu}}
  & {\small\texttt{michael.zhao@doordash.com}}
  & {\small\texttt{sinan@mit.edu}}
\end{tabular}
\end{center}

\footnotetext[1]{These authors contributed equally to this work.}
\footnotetext[2]{Corresponding author.}

\vspace{1.1em}
\begin{abstract}
\noindent
Using advertising campaign data from a large US-based mobile game developer, the authors study a global advertising shutoff in the context of mobile app install ads. Contrary to prior studies in search advertising---which reveal a major over-attribution problem where paid advertising takes credit for organic traffic that would have occurred otherwise---this study shows the opposite: paid ads generate positive spillovers to organic installs. Event study analysis shows that the shutoff decreased organic installs by 20--30\%. Fixed-effects panel models estimated on longer-term data find that every \$100 spent is associated with 32 paid installs and 2.2 organic installs, highly consistent with the event study estimates. Further analysis strongly suggests that this positive paid-to-organic spillover operates through a ranking mechanism: paid installs boost app store category rankings, thereby increasing organic visibility. Combining campaign and ranking data, the authors find that (1) ad spend has a statistically and economically significant relationship with store rankings; and (2) the relationship between organic installs and ad spend disappears once these rankings are factored in, indicating that they absorb the relationship. These findings demonstrate that mobile app install ads are more effective than paid install metrics alone indicate, implying that developers may systematically underinvest in marketing.

\end{abstract}

\clearpage
\doublespacing
\pagenumbering{arabic}
\setcounter{page}{1}

\label{section_intro}

Prior studies of high-profile ad shutoff experiments in paid search find that ads cannibalize organic search traffic for well-known brands, with organic recovering anywhere from 37\% to 99.5\% of paid effects \citep{blake2015consumer, Coviello2017large, golden2017effects, simonov2019competitive}. Fundamentally, these results point toward an underlying over-attribution problem where paid ads are taking credit for organic traffic that would have occurred otherwise \citep{berman2018attribution, li2014attributing}. Whether this issue extends to other digital environments is an open question. Consumers often use search navigationally (e.g., typing a brand name to reach its website), so paid and organic clicks are more likely to compete for the same underlying intent \citep{golden2017effects, simonov2018competition}.

We study this question in the context of mobile app install ads. Mobile apps have become a major consumer medium, with U.S. adults spending about 4 hours per day on mobile devices \citep{eMarketer2024time} and global app-store consumer spending reaching \$150 billion in 2024 \citep{sensortower2025}. In this environment, developers compete primarily through paid user acquisition.\footnote{Throughout, we use ``developers'' to refer to both app makers and app publishers. Our usage of ``publishers'' will refer to advertising publishers and platforms such as Google, Facebook, and Vungle, unless explicitly noted otherwise.} Millions of apps compete for visibility, with thousands released each day, while global app marketing spend reached \$109 billion in 2025 \citep{appsflyer2025}. Measuring the returns to that spending therefore matters, especially for firms operating on thin margins in markets with highly skewed consumer spending. However, industry standard last-touch / click attribution credits only the final ad touchpoint before conversion \citep{gordon2021inefficiencies}. As search is a major user acquisition channel, a similar over-attribution problem might also be present in this context. However, app store ranking lists and personalized recommendations offer additional discovery channels. Field experiments can address these biases \citep{lewis2015measuring, gordon2019comparison, gordon2022nonexperimental, gordon2023predictive}, though they remain costly and have historically been available mainly to large advertisers.

In this study, we leverage the nearly 2 years of advertising campaign data of a major US-based mobile game developer we will refer to as GameSpace. During this time period, GameSpace conducted a global advertising spend shutoff covering search, social, display, and in-app advertising. Using an event study around this global ad shutoff, we find that when the developer halted ad spending, contrary to findings in previous studies, organic installs dropped by 20--30\%, depending on bandwidth. To more precisely estimate effect sizes by dollar amounts, we also analyzed several fixed-effects panel models. We found that every \$100 of ad spend was positively associated with 32.3 paid and 2.2 organic installs. Lagged fixed-effects specifications suggest modest next-day paid carryover (approximately 2.8 installs per \$100, $p \approx 0.05$, see Web Appendix~A), consistent with delayed attribution, while lagged effects on organic installs are not statistically significant. These contemporaneous associations are qualitatively consistent with our event study findings.

By combining our advertising campaign data with app store ranking data, we find that ad spend is very strongly associated with app store rankings: during the shutoff, our event study analysis finds a 2.3--3.3-fold deterioration in raw chart position, corroborated by fixed-effects panel analysis. More crucially, we also find that the relationship between ad spend and organic installs disappears once we control for app ranking, indicating that rankings are mediating the effect of ad spend. These results strongly suggest that the positive paid-to-organic spillovers in our context are the result of an underlying ranking mechanism: paid advertising drives up installs, which then drive up app ranks, which leads to more downstream organic discovery and installs. 

While our work relies on app store ranking data, this ranking mechanism potentially extends to other forms of algorithmic ranking such as personalization and search as well. If the model training of such systems doesn't distinguish between paid and organic signals, then a similar loop to that described above occurs: paid advertising drives up installs, models train on this data and subsequently score apps higher, which leads to more downstream organic discovery and installs.

Our study contributes to both academic research and advertising practice. Our study provides empirical evidence of positive paid-to-organic spillovers in platform discovery environments, contrasting with the cannibalization and substitution effects widely documented in paid search. By identifying app store ranking as the key mediating mechanism, we highlight a critical methodological challenge: because rankings are determined at the platform level, localized advertising changes can spill over globally, violating SUTVA. Standard incrementality tests---which withhold ad spend from holdout users---may therefore suffer systematic downward bias by suppressing the counterfactual ranking lift that would otherwise drive organic installs. Taken together, our findings suggest that mobile app advertising is more effective than standard attributed install numbers imply, indicating that mobile advertisers may be systematically underinvesting in marketing.

\section*{Related Literature} \label{section_lit}

\subsection*{Advertising Shutoff Experiments} \label{section_lit_ad_shutoff}

Before advertising shutoff experiments, research on search advertising examined the impact of organic and sponsored search on ad performance. For example, \cite{ghose2009empirical} analyzed six months of panel data on search metrics across numerous keywords, revealing a direct correlation between conversion rates and search result rankings: the higher the rank, the better the conversion rates. Moreover, \cite{yang2010analyzing}, through an \textit{ad pulse} experiment, substantiated that paid search advertisements significantly enhance click-through rates and revenue.

In their pioneering study using a shutoff in ad spend, \cite{blake2015consumer} (henceforth BNT) produced strong evidence that paid search advertising substitutes for organic traffic. Using a series of large-scale field experiments conducted at eBay, they showed that when branded search ads were suspended, most of the lost advertising traffic simply shifted to organic search traffic. One interesting heterogeneous effect was that non-branded search ads had a positive effect on new or infrequent consumers. On average, however, returns on advertising were ultimately negative since most ads were delivered to frequent consumers who, at least in the short run, were not affected by ads. 

Later, \cite{Coviello2017large} were concerned with the generalizability of BNT's findings given that eBay was particularly well-known at the time. They replicated an experiment at Edmunds.com similar to BNT's at eBay. They found the same substitution effect but at a far smaller magnitude: less than half of paid search traffic was recovered through organic search. Such an effect suggests that paid search may provide a positive return, but it still points to substitution of paid advertising with organic traffic.

\cite{simonov2019competitive} and \cite{golden2017effects} further confirmed \cite{Coviello2017large}'s results. Although both studies were primarily focused on understanding how a firm's paid search advertising impacts their competitors' outcomes, their direct results show that paid search is indeed effective but that it crowds out organic traffic to some degree. In the case of \cite{golden2017effects}, they noted that paid advertising is approximately 63\% efficient, implying that 37\% of paid advertising is substituted by organic traffic. Similarly, \cite{simonov2019competitive} found that paid search ads cannibalized about 37.8\% of a brand's organic search traffic. Other studies also examine the impact of search result competition on conversion performance, noting attenuation in performance with increased competition \citep{agarwal2016organic, bhattacharya2021competitive}.

Our work builds on this existing literature in several ways. First, our work is not limited to paid search advertising. While paid search is still a dominant channel in digital advertising, other channels continue to gain market share each year, reaching 61\% of all digital ad spend in 2019 \citep{marinsoftware2019}. Second, unlike prior ad shutoff experiments, our results uncover evidence of positive organic spillovers from paid advertising rather than cannibalization. While positive digital advertising spillovers have been documented in other contexts---such as competitive spillovers to non-advertised brands \citep{sahni2016advertising}, online-to-offline spillovers \citep{kalyanam2018search}, and consumer native ad spillovers \citep{sahni2020sponsorship}---our results demonstrate a positive organic spillover back to the focal firm's own digital channel under a global ad shutoff. Third, our work brings ad shutoff experiments to the mobile channel. As mobile use and mobile advertising are both growing at unprecedented rates, understanding the extent to which paid advertising generates organic spillovers on the fastest-growing devices is essential. In a sense, our work brings the ad shutoff experiment literature from the desktop era to the mobile era.

\subsection*{App Store Rankings and Discovery} \label{section_lit_rankings}

Despite growing interest in mobile advertising, surprisingly little work evaluates mobile app install advertising using field experiments. Existing research has focused mainly on machine learning for bidding and targeting \citep{ma2016app2vec, bhamidipati2017large, sahu2018managing}. Our work is among the first to assess install-ad effectiveness using a large-scale spending shutoff experiment.

A related stream of work studies how published app store rankings shape consumer demand and firm strategy. \cite{carare2012bestseller} uses daily rank data from Apple's App Store to show that today's bestseller rank causally raises tomorrow's willingness to pay by roughly \$4.50 for a top-ranked app, with effects declining down the top-100 list. \cite{deng2023freemium} study freemium launches in the App Store and attribute part of the spillover from a free to a paid version to enhanced app discovery through top-chart visibility, using category ranking data as a key outcome. These studies establish that rankings matter for demand and discovery, but neither examines whether paid install advertising moves category rankings or uses an ad shutoff to identify the advertising-to-ranking link.

A smaller literature asks whether firms can strategically manipulate chart position. \cite{li2016buyingdownloads} model and estimate how developers ``buy downloads'' to climb top-app lists, finding that \$100 of purchased downloads improves an app's ranking by roughly 2.2\%. \cite{dover2015rankamplification} study Facebook advertising for a paid smartphone app sold on the App Store and document a ``rank amplification'' mechanism where advertising boosts sales, which improves the published sales rank, thereby driving subsequent sales. While their study provides an elegant modeling and simulation-based analysis for a single paid app, our work differs in several critical dimensions. First, we examine user acquisition (organic and paid installs) in the context of free-to-play apps on category-level free charts rather than purchase decisions on paid bestseller charts. Second, while \cite{dover2015rankamplification} rely on observational data and simulation, we leverage a massive, clean, and exogenous multi-platform ad shutoff to provide direct, causal, and experimental evidence of this mechanism. Finally, we trace and quantify these links (from advertising to category rank, and from category rank to organic installs) using a combination of event studies, lagged panel regressions, and fixed-effects mediation analysis. To our knowledge, our study is the first to provide comprehensive empirical evidence of this ranking-mediated mechanism using large-scale field experimental variation. This platform-mediated discovery process represents a structural departure from consumer-level spillover mechanisms documented in other digital environments \citep{sahni2020sponsorship}, where spillovers operate through consumer cognitive processing and subsequent recall. In contrast, the spillover we document operates through a system-level algorithmic feedback loop: paid advertising drives a surge in install velocity, which alters the public ranking state and subsequent organic discovery.

\subsection*{Attribution Models}  \label{section_attribution}

The study of attribution models in digital advertising is key to understanding their impact on online marketing strategies and budget allocation \citep{li2016attribution, danaher2018delusion, danaher2020advertising, lewis2025amazon}. In particular, \cite{berman2018attribution} highlighted the drawbacks of last-touch attribution in multi-publisher online advertising, which can lead to excessive ad exposures and distorted market pricing. However, standard last-touch models often fail to capture indirect spillovers across channels \citep{gordon2021inefficiencies}.

While attribution models were not our primary focus, our work relates the literature on attribution models to the literature on ad effectiveness, in the context of mobile apps. First, ad effectiveness is estimated through attribution models. While previous studies show that attribution can overestimate ad effectiveness when search ads function as navigational substitutes, our work demonstrates how attribution can underestimate ad effectiveness in ranking-mediated mobile ecosystems where ads drive organic rank spillovers. Moreover, \cite{li2014attributing} show that last-touch attribution significantly underestimates the contribution of e-mails, display ads, and referrals to conversions. They demonstrated that some customers saw ads on one channel and searched on a search engine as a navigational tool, as also shown by \cite{golden2017effects} and \cite{simonov2018competition}. We report exploratory platform-decomposed regressions in Web Appendix~C that are consistent with this broader attribution concern---last-touch models may understate non-search channels such as Facebook relative to search---though we cannot identify causal platform-level effects from our aggregate shutoff variation.

\section*{Data} \label{section_data}
\begin{figure}[!htbp]
\centering
{\includegraphics[width=\linewidth]{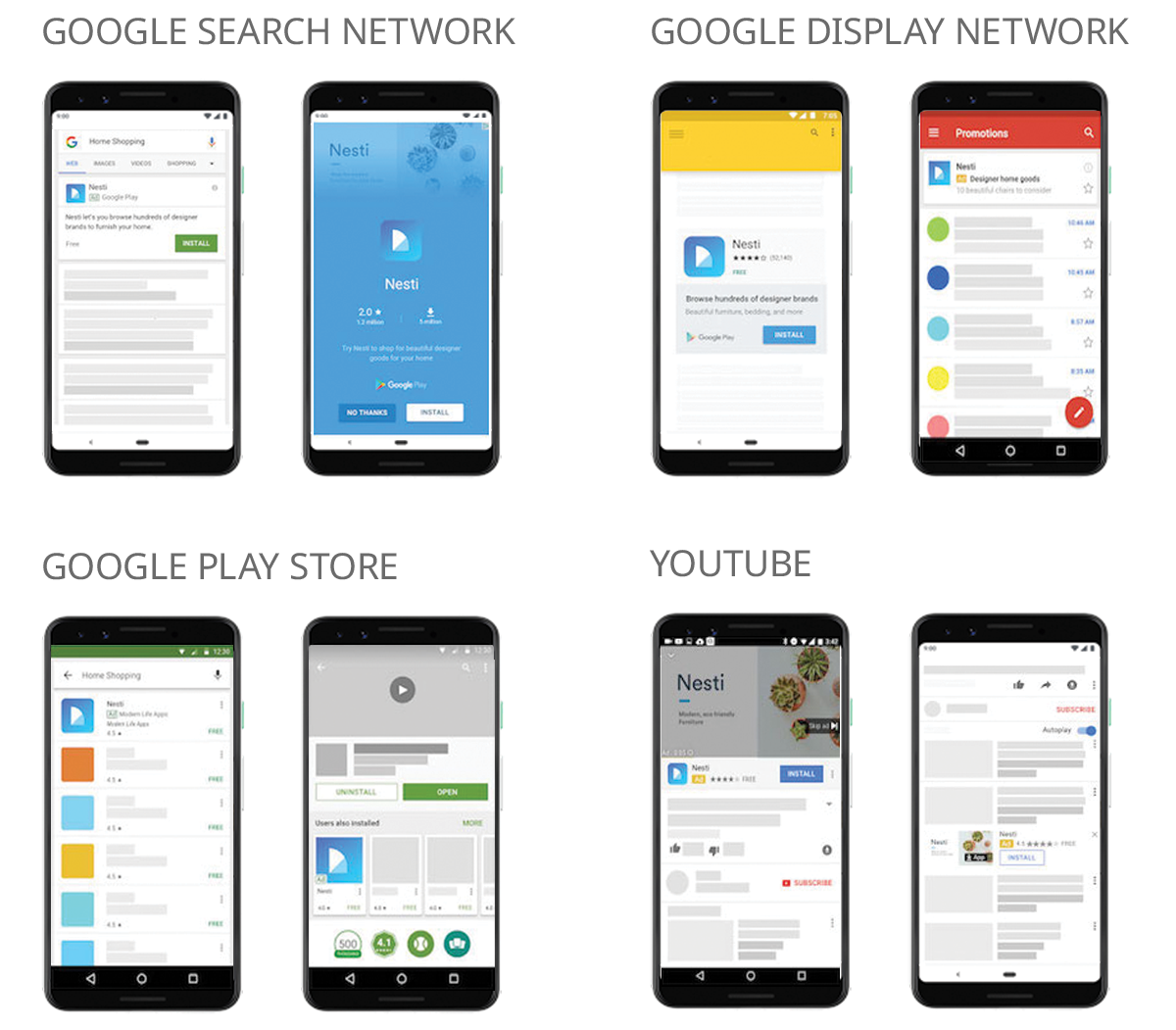}}
\caption{Examples of Mobile App Install Ads} \label{fig_ad_examples}
\end{figure}

Mobile app install ads are advertisements designed to drive installs of a mobile app. Although they can appear across the entire spectrum of digital channels (\textit{e.g.}, search, social media, display, in-app, video), they generally link to an app's listing in an app store to allow consumers to install the app from the ad directly. Moreover, they are generally mobile-only to make the app installation process as frictionless as possible. Figure \ref{fig_ad_examples} illustrates four examples of such ads. 

We conducted our analysis on the historical advertising campaign data of a major game developer, which we refer to as GameSpace. These data were provided to us through our collaboration with a US-based startup, which we call AdTech, that manages and optimizes digital advertising spend on behalf of GameSpace and other clients. The data track the daily spend, impressions, clicks, and installs for all the digital advertising campaigns set up by GameSpace across 85 different ad publishers for 500 days in 2018 and 2019 (calendar dates are withheld due to privacy concerns). Paid installs are installs that occur as a result of a user clicking on a paid advertisement. We used last-touch attribution, crediting the last ad impression with the install rather than any previous impressions. Organic installs are installs that occur without a user clicking on any paid advertisements. Overall, organic installs account for just over 40\% of all installs. To provide a sense of the scope of this data, over 100 billion impressions were served worldwide during our time period.

To investigate the underlying mechanism of organic lift, we also collect daily category ranking history from Appfigures, a leading third-party mobile market intelligence platform, for each of our six games across both iOS and Android platforms in the United States storefront. The ranking data cover the same 500-day panel window as the advertising data, allowing for a complete merge with our primary advertising panel. We track positions on each app's primary category-level Free chart: iOS Puzzle Free (category ID 7012) and Android Games Free (category ID 43). These represent the main storefront lists where the games actively competed for organic visibility, and are the primary categories under which the developer's games are listed and where they rank highest on average (matching the specific puzzle subcategory on iOS and the broader games category on Android). Appfigures does not report a rank on days when an app falls off the tracked category chart. Since standard app store chart rankings are numerically inverted (where 1 represents the top chart position and larger numbers represent worse rankings), we log-transform the raw chart position, denoting the result $\log\text{Rank}_{ijt}$ for app $i$ on operating system $j$ on date $t$. Higher values of $\log\text{Rank}_{ijt}$ therefore indicate worse chart positions. Category ranks are also heavy-tailed: moving from rank~5 to rank~10 conveys much more visibility than moving from rank~500 to rank~505. A one-unit increase in $\log\text{Rank}_{ijt}$ therefore corresponds to an $e$-fold deterioration in raw chart position (e.g., rank~20 to rank~$\approx$55). Web Appendix~B replicates the rank-mechanism results using $-\log(\text{Rank}_{ijt})$, $-\text{Rank}_{ijt}$, and $1/\text{Rank}_{ijt}$, with qualitatively unchanged conclusions.

For our analysis, we aggregated this historical campaign data by summing installs and spending for all campaigns for each app and operating system (OS) combination at a daily level (so each observation of the data is uniquely identified by an app, OS, and date combination). We restricted our analysis to 6 particular mobile apps out of GameSpace's larger portfolio, leading to a dataset with 6000 observations (6 apps $\times$ 2 OS $\times$ 500 dates). We focused specifically on these 6 since they had the most complete and active advertising activity throughout our data. Furthermore, all these apps were relatively mature, with the youngest being released over 6 months before the beginning of our observation period, allowing us to avoid any inflation in organic install numbers due to media mentions. Despite all apps being relatively mature and having active advertising activity, there was still some heterogeneity with the most popular app garnering around 4 times the spend and installs of the least popular app.

The estimation sample excludes dates with app-store featuring spikes, which would otherwise confound organic-install dynamics, leaving 5829 app-OS-date observations. For rank-mechanism analyses, our main specifications use observed Appfigures chart positions wherever reported (approximately 93\% of these observations); off-chart days are omitted because Appfigures does not report them. Web Appendix~B shows that the rank-mechanism results are robust to alternatively coding off-chart days as rank~401, just below Appfigures' default top-400 cutoff.

\begin{figure}[!htbp]
\centering
{\includegraphics[width=\linewidth]{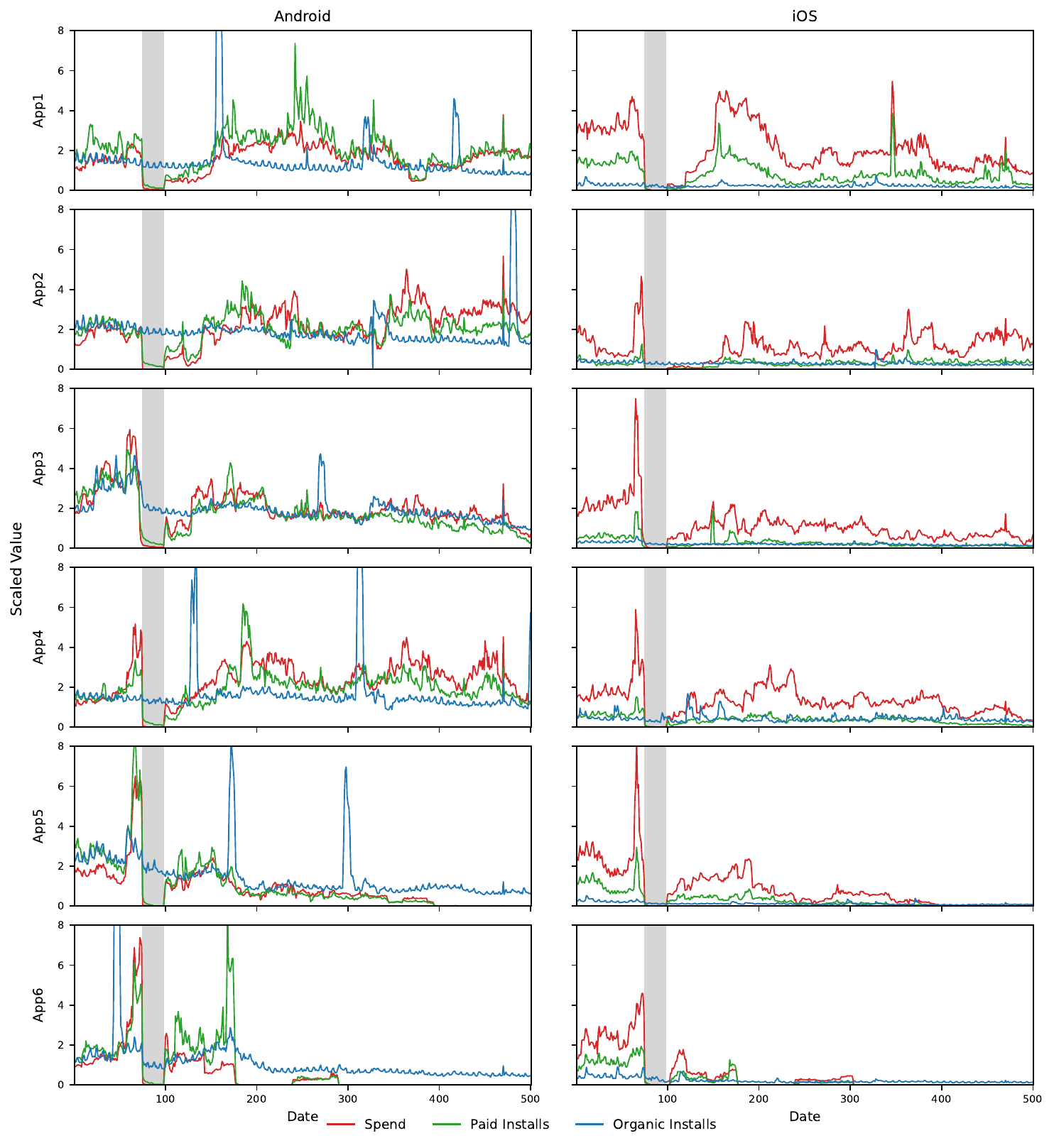}}
\caption{\small Scaled Time Series. \textmd{The scaled time series for spend, paid installs, and organic installs by app and OS. Each row denotes an app and each column denotes an OS. Scaled values are produced by dividing the app-OS-date summed numbers by the standard deviation of that metric across both OSes. This means that each metric is scaled separately within app--OS pairs. Note that there are several instances of unusually high organic installs. These spikes occurred when an app was somehow featured in the respective app store.} 
} \label{fig_time_series}
\end{figure}

To visualize the data, we plotted the scaled values for spend, paid installs, and organic installs for the 12 different app-OS combinations in Figure \ref{fig_time_series}. Scaled values were computed by dividing each series by the standard deviation of that metric for that particular app. Thus, the scaling factor is preserved between device OS (\textit{e.g.}, App1 Android spend and App1 iOS spend are divided by the same number) but will differ across metric and app (\textit{e.g.}, App1 iOS spend, App1 iOS paid installs, and App2 iOS spend are all divided by different numbers).

\begin{figure}[!htbp]
\centering
{\includegraphics[width=\linewidth]{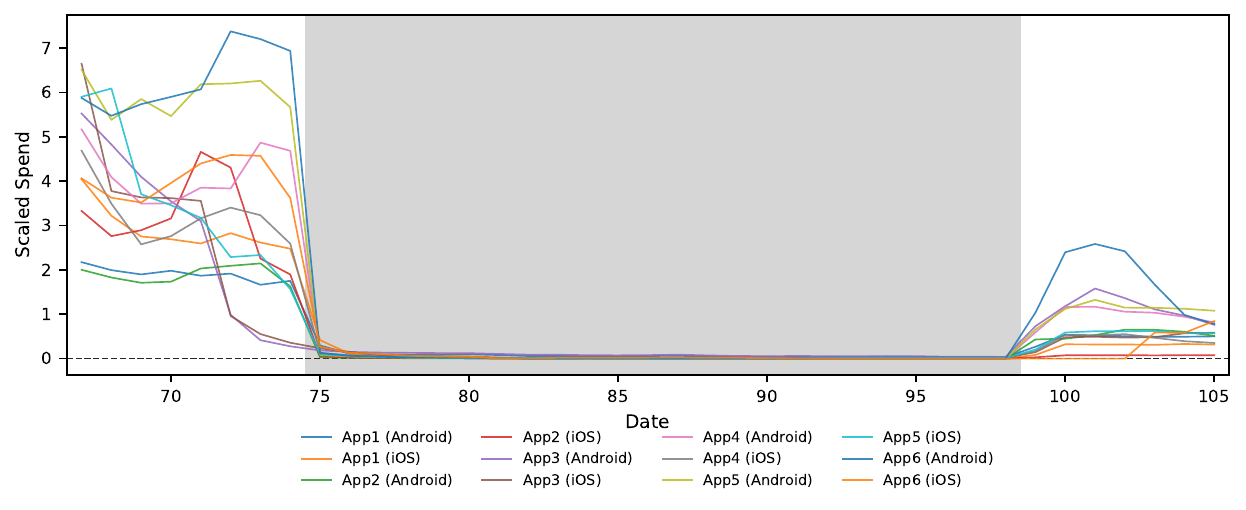}}
\caption{\small Scaled Spend during Shutoff. \textmd{Scaled spend on days 67 through 105 for all of the 12 different app-OS combinations we analyze in this study around the shutoff period, highlighted in gray shading.}} \label{fig_spend_zoomed}
\end{figure}

\begin{figure}[!htbp]
\centering
{\includegraphics[width=\linewidth]{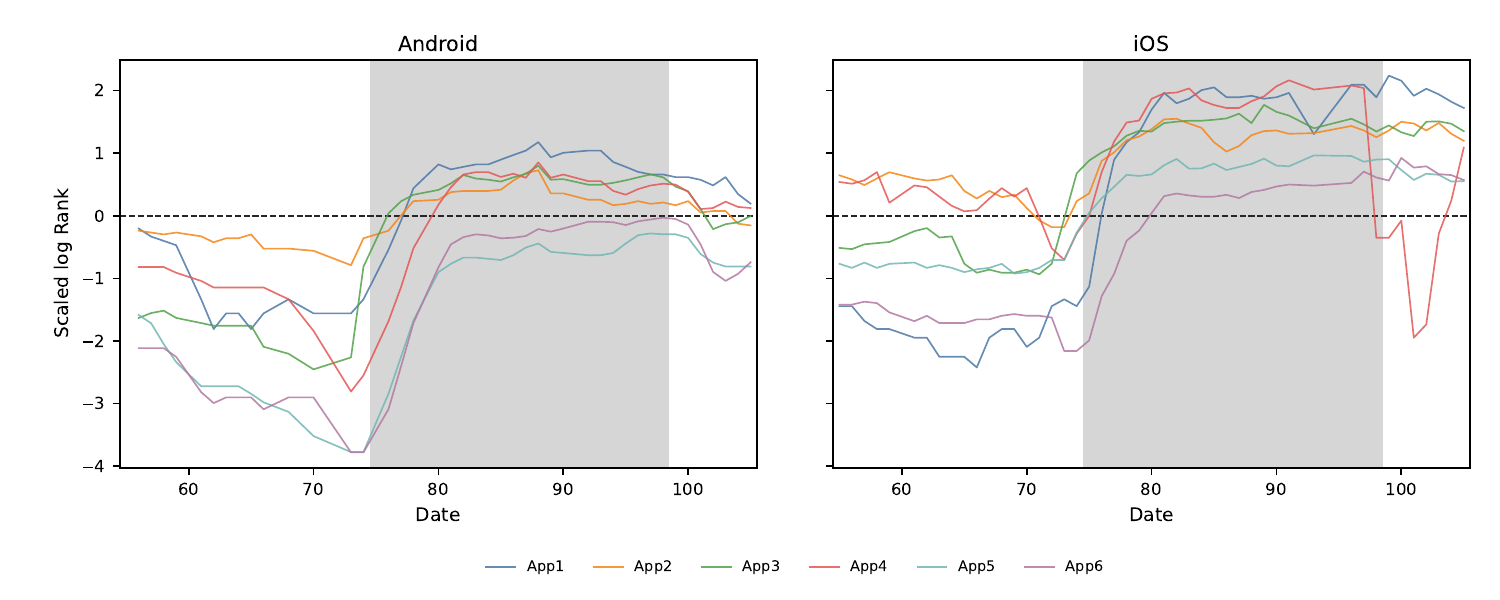}}
\caption{\small Scaled Category Rank during Shutoff. \textmd{Demeaned and SD-scaled $\log\text{Rank}$ on days 55 through 105 around the global shutoff (gray shading). Left: Android; right: iOS. Series are demeaned and scaled within each app across both platforms so absolute chart positions are not identifiable. Higher values indicate worse chart positions; the dashed line marks each app's sample mean.}} \label{fig_rank_zoomed}
\end{figure}

The key to our identification strategy is a global ad shutoff experiment similar to the branded search shutoff experiments conducted by BNT at eBay. GameSpace implemented a coordinated shutoff across all apps and platforms between days 75 and 98 of our data collection (highlighted in dark gray shading in Figure \ref{fig_time_series}). Figure \ref{fig_spend_zoomed} examines this period in more detail to give a better sense of the underlying dynamics. Specifically, GameSpace stopped adding additional spending to their accounts several days earlier. As many campaigns still had positive balances, it took several days for the remaining budget to be exhausted. Looking closely at Figure \ref{fig_spend_zoomed}, we can see some series never quite reach 0 (indicated by the black dashed line) during the shutoff for this reason, but advertising spend is practically zero during this period. We also note the heightened amount of spending, at least for app-OS pairs in the weeks leading up to the global shutoff. This was due to GameSpace moving its spending forward, using what it would have spent during the shutoff in the weeks before. Figure \ref{fig_rank_zoomed} displays the corresponding raw, scaled category rank series around the shutoff window. Ranks are log-transformed, then demeaned and SD-scaled within each app (pooling Android and iOS) so absolute chart positions are not identifiable. Higher values indicate worse chart positions (same direction as $\log\text{Rank}_{ijt}$ in our analysis). Two additional partial shutoffs for individual apps (App5 from day 395 and App6 from days 177--239) provide supplementary context but are not our primary identification strategy. All apps were active during and after the 500 days that we observed.

\section*{Empirical Strategy and Model Specifications} \label{section_model}

We employed two main empirical methods to analyze the paid ad shutoff experiments across GameSpace's six different apps: an event study and fixed effects panel regressions. We describe the details of each method and the corresponding model specifications below.

\subsection*{Event Study} \label{section_model_es}
In our event study estimation, we treat the global ad shutoff as the event of interest and estimate outcome changes in a local window around the shutoff date \citep{hausman2018regression}. Our event date is the start of the global shutoff on day 75. While ad spend reactivation theoretically serves as an additional event, we excluded it from our analysis since reactivation was gradual (Figure \ref{fig_spend_zoomed}). We estimate local constant and local linear specifications:
\begin{align}
\log(Y_{ijt}) &= \delta D_{ijt} + \alpha_{ij} + \omega_t + \epsilon_{ijt}\label{eq_local_constant}\\
\log(Y_{ijt}) &= \delta D_{ijt} + \gamma_1 (t - c) + \gamma_2 D_{ijt} (t - c) +\alpha_{ij} + \omega_t + \epsilon_{ijt} \label{eq_local_linear}
\end{align}
where $Y_{ijt}$ denotes organic, paid, or total installs for app $i$ on OS $j$ at date $t$; $D_{ijt}$ is an indicator equal to one after the shutoff cutoff $c$; $\alpha_{ij}$ are app-OS fixed effects; $\omega_t$ are day-of-week fixed effects; and $\epsilon_{ijt}$ is the error term. We log-transform install counts to interpret $\delta$ as an approximate proportional change, which is appropriate given substantial heterogeneity in app scale. The coefficient $\delta$ captures the discontinuity in installs at the shutoff. Poisson robustness checks appear in the Web Appendix. Because time is the running variable, calendar date fixed effects would be collinear with $D_{ijt}$ during the estimation window. For Equation \ref{eq_local_linear}, $(t - c)$ is the difference in days between date $t$ and cutoff $c$. The slope coefficients $\gamma_1$ and $\gamma_2$ are not the focus of our analysis.

We purposefully limited ourselves to the simpler event study specifications since the more complex specifications run the risk of overfitting. Overfitting can be especially problematic since the decrease in spending, while rather drastic, is still not perfectly sharp, thus extending the outcome shift across multiple days. As such, higher-order polynomial terms, if used, would more likely capture the underlying shift in spending, rather than the true nonlinearity of interest. For instance, as seen in Figure \ref{fig_spend_zoomed}, the two series corresponding to App3 already attained a much lower level of spend 3 days before the cutoff compared to the other apps. Though this is the most prominent example, many other apps already had declining ad spend in the days before $c$. Furthermore, persistence effects are a concern: a consumer may install an app today due to ad exposure yesterday or the day before. While such effects will eventually decay, they will also extend the outcome shift across multiple days. 

\subsection*{Fixed Effects Panel Regression} \label{section_model_fe}

\noindent\textit{Aggregate Spend.}\par
\noindent Although the event study identifies the causal effect of a shutoff in ad spend, in practice, firms adjust spend continuously rather than shutting it off entirely. To examine how day-to-day changes in ad spend relate to installs, we used the panel regression in Equation \ref{eq_spend}:
\begin{equation}
Y_{ijt} = \beta \text{spend}_{ijt} + \pi_{ijt}  + \epsilon_{ijt} \label{eq_spend}
\end{equation}

Here, our main variable of interest is $\text{spend}_{ijt}$, the number of dollars spent on advertising app $i$ for OS $j$ on date $t$ across all publishers and campaigns. For this specification, we did not log transform our variables because we expect the underlying relationship between spending and installs (organic, paid, or total), is linear rather than multiplicative. We avoided using date fixed effects as date-level shocks are likely multiplicative rather than additive because calendar shocks (e.g., weekends) are likely multiplicative across apps of different scale. Instead, we employed an app-OS-time fixed effect $\pi_{ijt}$ that consists of the interactions between the app-OS fixed effects $\alpha_{ij}$ and day-of-week fixed effects $\omega_t$ as well as the interactions between app-OS fixed effects $\alpha_{ij}$ and week fixed effects $\psi_{t}$. More formally $\pi_{ijt} = \alpha_{ij} \times \omega_t + \alpha_{ij} \times \psi_t$. Thus, the effect $\beta$ captures the deviations relative to the average number of installs on an app-os-day-of-week and an app-os-week level. Exploratory platform-decomposed specifications appear in Web Appendix~C.

\subsection*{Rankings and Mediation Specifications} \label{section_model_mechanism}

To investigate the underlying mechanism of the organic lift, we formally specify the models used to test the ranking mechanism, using $\log\text{Rank}_{ijt}$.
\noindent\textit{Event Study on App Store Rankings.}\par
\noindent To test whether paid ad spend causally affects category rankings, we replicate the event study design from Equations~\ref{eq_local_constant} and \ref{eq_local_linear} but substitute $\log\text{Rank}_{ijt}$ as the dependent variable:
\begin{align}
\log\text{Rank}_{ijt} &= \delta D_{ijt} + \alpha_{ij} + \omega_t + \epsilon_{ijt} \label{eq_rank_rdit_constant}\\
\log\text{Rank}_{ijt} &= \delta D_{ijt} + \gamma_1 (t - c) + \gamma_2 D_{ijt} (t - c) + \alpha_{ij} + \omega_t + \epsilon_{ijt} \label{eq_rank_rdit_linear}
\end{align}
where $D_{ijt}$ is the post-shutoff indicator, $\alpha_{ij}$ are app-OS fixed effects, $\omega_t$ are day-of-week fixed effects, and $(t - c)$ is the running variable of days relative to the cutoff.

\noindent\textit{Lagged Effects of Advertising on Store Rankings.}\par
\noindent To examine whether yesterday's advertising spend predicts today's store ranking, we estimate the following panel regression with lagged spend:
\begin{equation}
\log\text{Rank}_{ijt} = \theta \text{spend}_{ij(t-1)} + \pi_{ijt} + \epsilon_{ijt} \label{eq_rank_temporal}
\end{equation}
where $\pi_{ijt} = \alpha_{ij} \times \omega_t + \alpha_{ij} \times \psi_t$ represents the app-OS-time fixed effects defined in Equation~\ref{eq_spend}.

\noindent\textit{Mediation Model Specification.}\par
\noindent To evaluate whether category rankings mediate the relationship between contemporaneous spend and installs, we estimate parallel fixed-effects specifications containing both variables as predictors:
\begin{equation}
Y_{ijt} = \beta_1 \text{spend}_{ijt} + \beta_2 \log\text{Rank}_{ijt} + \pi_{ijt} + \epsilon_{ijt} \label{eq_rank_mediation}
\end{equation}
where $Y_{ijt}$ represents organic or paid installs, $\text{spend}_{ijt}$ is contemporaneous spend across all ad campaigns, $\log\text{Rank}_{ijt}$ is the log-transformed app store category chart ranking, and $\pi_{ijt}$ is the app-OS-time fixed effects. Under the ranking mechanism, adding store ranking to the model should absorb the association between contemporaneous spend and organic installs, while paid installs (which are driven mainly by contemporaneous ad delivery) should remain strongly associated with spend.

\section*{Results} \label{section_results}

We organize the presentation of our results into three subsections. The first two outline event study and fixed effects results. The third examines app store rankings as a mechanism for organic lift. Local linear event study estimates, Poisson robustness checks, lagged fixed-effects specifications, bandwidth-sensitivity figures, and exploratory platform-decomposed regressions appear in the Web Appendix. We used the event study model to analyze the experimental results and the fixed effects regression to estimate the precise effect of ad spend per dollar on organic installs. In the event study results, we report local constant estimates (Equation \ref{eq_local_constant}) for several different choices of bandwidth. In the fixed effects results, we report our panel regression estimates (Equation \ref{eq_spend}). To account for unobserved events like app or OS updates, we report all our results with app-clustered standard errors. Our analysis was conducted in Python 3.12 using \texttt{pyfixest} \citep{fixest} and \texttt{polars}.

\subsection*{Event Study} \label{section_results_shutoff_es}

\begin{figure}[!htbp]
\centering
{\includegraphics[width=\linewidth]{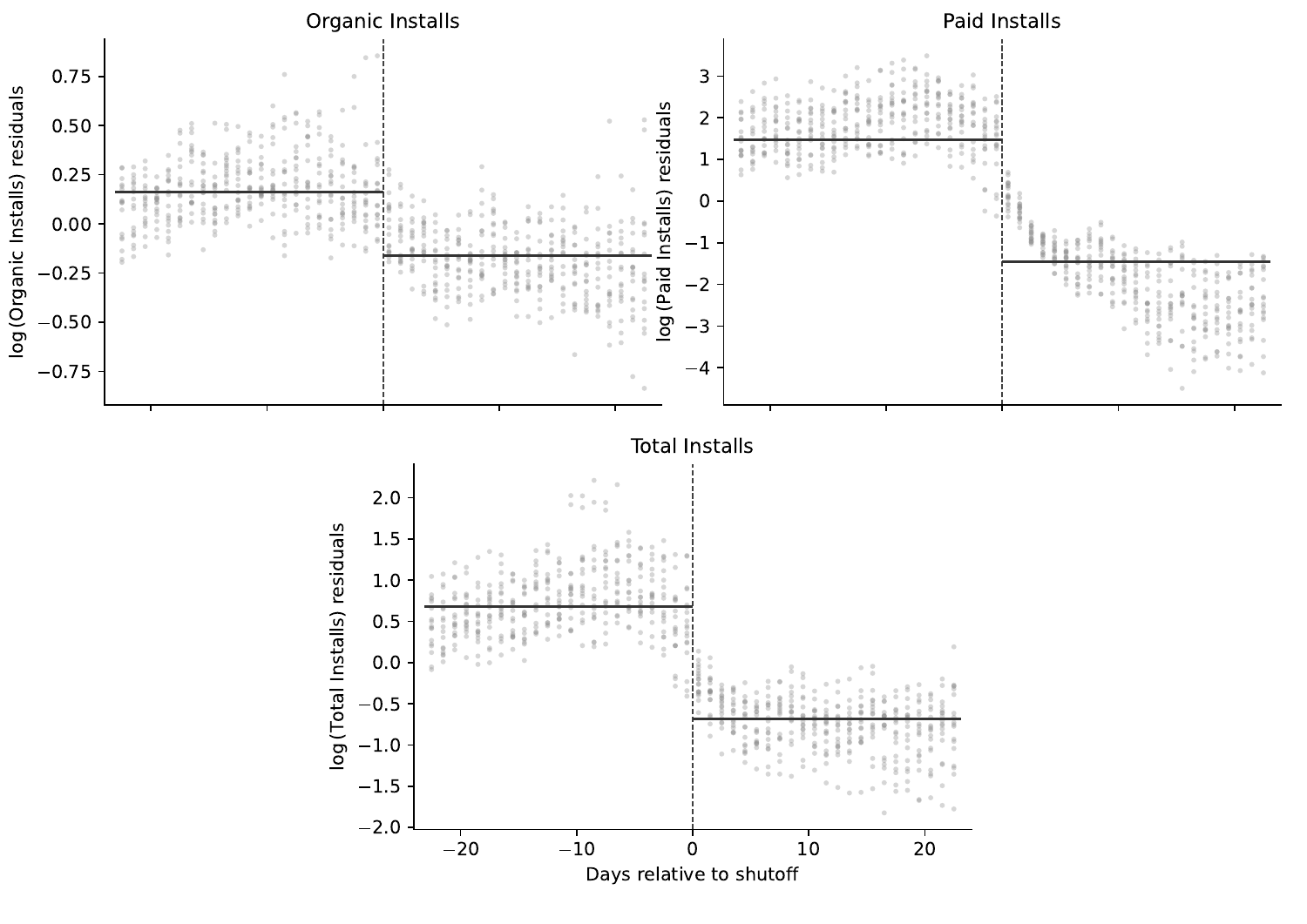}}
\caption{\small Event Study. \textmd{Residuals after controlling for app-OS and day-of-week fixed effects over an 11-day window around the global shutoff. Organic installs (top left), paid installs (top right), and total installs (bottom center). The line shows the local constant fit at an 11-day bandwidth. Points are jittered for visualization.}} \label{fig_es}
\end{figure}

We estimated the impact of the ad spend shutoff on organic installs using the event study model described above. We plotted the local constant regression fit from Equation \ref{eq_local_constant} in Figure \ref{fig_es} at an 11-day bandwidth for each of the three outcomes. The points on the plot are formed using the residuals of $\log(Y_{ijt})$ after controlling for the app-OS fixed effects ($\alpha_{ij}$) and day-of-week fixed effects ($\omega_t$). We report the estimated discontinuities and app-clustered standard errors for representative bandwidth choices (7, 11, and 23 days) in Table \ref{table_es}. Web Appendix~A reports the full local constant grid from 3 to 23 days and shows that the visual fit is stable across bandwidths.

\begin{table}[H]
\centering
\begin{tableresize}
\begin{threeparttable}
\begin{tabular}{@{\extracolsep{5pt}}rccc}
\\[-1.8ex]\hline
\hline \\[-1.8ex]
& \multicolumn{3}{c}{\textit{Dependent variable:}} \\
\cline{2-4}
\\[-1.8ex] Bandwidth & \multicolumn{1}{c}{Organic Installs} & \multicolumn{1}{c}{Paid Installs} & \multicolumn{1}{c}{Total Installs}  \\
\hline \\[-1.8ex]
7 days       & $-0.305$ & $-2.732$ & $-1.311$\\
            & (0.061; p = .004) & (0.203; p = .001) & (0.179; p = .001) \\ \\
11 days       & $-0.319$ & $-3.013$ & $-1.368$\\
            & (0.066; p = .005) & (0.235; p = .001) & (0.189; p = .001) \\ \\
23 days       & $-0.359$ & $-3.709$ & $-1.460$\\
            & (0.081; p = .007) & (0.311; p = .001) & (0.208; p = .001) \\ \\
\hline \\[-1.8ex]
Obs               & 825 & 825 & 825 \\
\hline
\hline
\end{tabular}

\caption{Event Study Estimates of Ad Shutoff Effects}  \label{table_es}
\end{threeparttable}
\end{tableresize}
\end{table}

Beginning with the results on organic installs, regardless of our choice of bandwidth, the estimated discontinuity is always negative and highly statistically significant. The estimated discontinuity becomes stable at approximately $-0.32$ after bandwidth reaches 11 days before and after the cutoff. The results for paid installs and total installs are also qualitatively similar.

Smaller bandwidths typically provide less biased estimates of the local average treatment effect (LATE) of interest. However, there are some concerns with following this general rule in our context. Recall that our treatment is a blunt approximation of the underlying effect of advertising spend which ramps up and ramps down in the weeks before the shutoff. Local linear fits in Web Appendix~A show that the pre-shutoff slope changes from negative to positive as bandwidth increases, so the smallest bandwidths here may not provide the best estimate of the counterfactuals of interest. Moreover, as we mentioned earlier, there were likely to be some persistent effects of advertising in the previous days that take time to fully decay. Local linear fits also show negative post-shutoff slopes across bandwidths, suggesting that the effects of advertising lasted beyond the shutoff and ramped down over time.

Taking the above concerns into account, we considered the medium-bandwidth Local Constant results to be the most credible estimates of a shutoff effect. The point estimates of the medium-bandwidth results suggest that shutting off spend decreases organic installs by about 27\%. Recall, however, that GameSpace shifted much of the spend forward before the global shutoff, thereby inflating the amount of spending that occurred before the shutoff. Thus, our event study estimates compare elevated pre-shutoff spend to near-zero spend during the shutoff.

\subsection*{Fixed Effects Regressions} \label{section_results_fe}

As mentioned earlier, the event study uses a blunt measure of spend (\textit{i.e.}, shutting off spend entirely) that only allows us to estimate extreme counterfactuals. Thus, in this section, we used a fixed effects model for a more fine-grained associational estimate of the relationship between ad spend and installs. Having established a negative discontinuity in organic installs at shutoff, this fixed effects panel regression does not have an explicit identification strategy aside from assuming that fixed effects control for unobserved confounders. We estimated Equation~\ref{eq_spend}. Results are in Table \ref{table_panel}. Coefficients and standard errors are multiplied by 100 (installs per \$100). Lagged specifications appear in Web Appendix~A.

\begin{table}[H]
\centering
\begin{tableresize}
\begin{threeparttable}
\begin{tabular}{@{\extracolsep{5pt}}lccc}
\\[-1.8ex]\hline
\hline \\[-1.8ex]
& \multicolumn{3}{c}{\textit{Dependent variable:}} \\
\cline{2-4}
\\[-1.8ex] & Organic Installs & Paid Installs & Total Installs \\
\hline \\[-1.8ex]
$\text{spend}_{ijt}$  & $2.177$ & $32.295$ & $34.472$\\
 & (3.042; p = .51) & (3.867; p = .001) & (6.638; p = .003)\\ \\
\hline \\[-1.8ex]
R$^2$            & 0.942 & 0.919 & 0.936 \\
Obs               & 5829           & 5829           & 5829           \\
\hline
\hline
\end{tabular}

\caption{Ad Spend Generates 2.2 Organic and 32.3 Paid Installs per \$100} \label{table_panel}
\end{threeparttable}
\end{tableresize}
\end{table}

In line with our prior ad shutoff results, we found that advertising spend has a positive impact on organic installs, again suggesting underestimation of ad effectiveness based on only attributed installs. Table~\ref{table_panel} indicates that every \$100 spent on advertising is associated with approximately 2.2 additional organic installs on the same day, though this aggregate coefficient is not statistically significant with our updated clustering.

In our estimation of paid installs, we naturally see a much larger effect on spend. Table~\ref{table_panel} shows that GameSpace obtained around 32.3 paid installs per \$100 spent on advertising. Lagged specifications in Web Appendix~A suggest modest next-day paid carryover: every \$100 spent is associated with approximately 29.7 paid installs on the same day and 2.8 paid installs the next day ($p \approx 0.05$), consistent with delayed attribution. Lagged spend coefficients for organic installs are not statistically significant.

Overall, our results suggest that advertising spend generates more installs than attributed paid installs alone would suggest. From the perspective of \textit{cost-per-install} (CPI), attributed paid installs imply a CPI of \$3.10 (\$100/32.295) from the paid installs estimate in Table~\ref{table_panel}. However, CPI decreases to \$2.90 (\$100/34.472) when we account for organic lift and use the total installs estimate in the same table. While the mechanism of the organic lift is explored in detail in the subsequent section, our results show that attribution modeling without experiments will underestimate the effectiveness of mobile app install ads.

\subsection*{Spillovers via App Store Rankings} \label{section_results_mechanism}

Mobile app marketplaces differ from traditional desktop search environments. In desktop search, paid ads and organic results sit side-by-side. Buying an ad can directly cannibalize the organic click. In mobile app ecosystems, however, platform design introduces a distinct \textit{ranking spillover}. Prior work shows that chart rankings causally affect app demand \citep{carare2012bestseller} and that advertising can amplify sales through published rank information \citep{dover2015rankamplification}; developers can also invest directly to climb top lists \citep{li2016buyingdownloads}. App stores (Apple App Store and Google Play) use rolling windows (e.g., 24--72 hours) of install velocity as the primary signal for computing ``Top Charts'' and category rankings. Paid advertising campaigns generate a surge of paid installs, which artificially inflates this install velocity. The improved ranking increases the app's organic visibility to browsing consumers, driving a subsequent surge of organic installs, a discovery channel related to, but distinct from, the freemium spillovers documented by \cite{deng2023freemium}.

\begin{figure}[!htbp]
\centering
{\includegraphics[width=\linewidth]{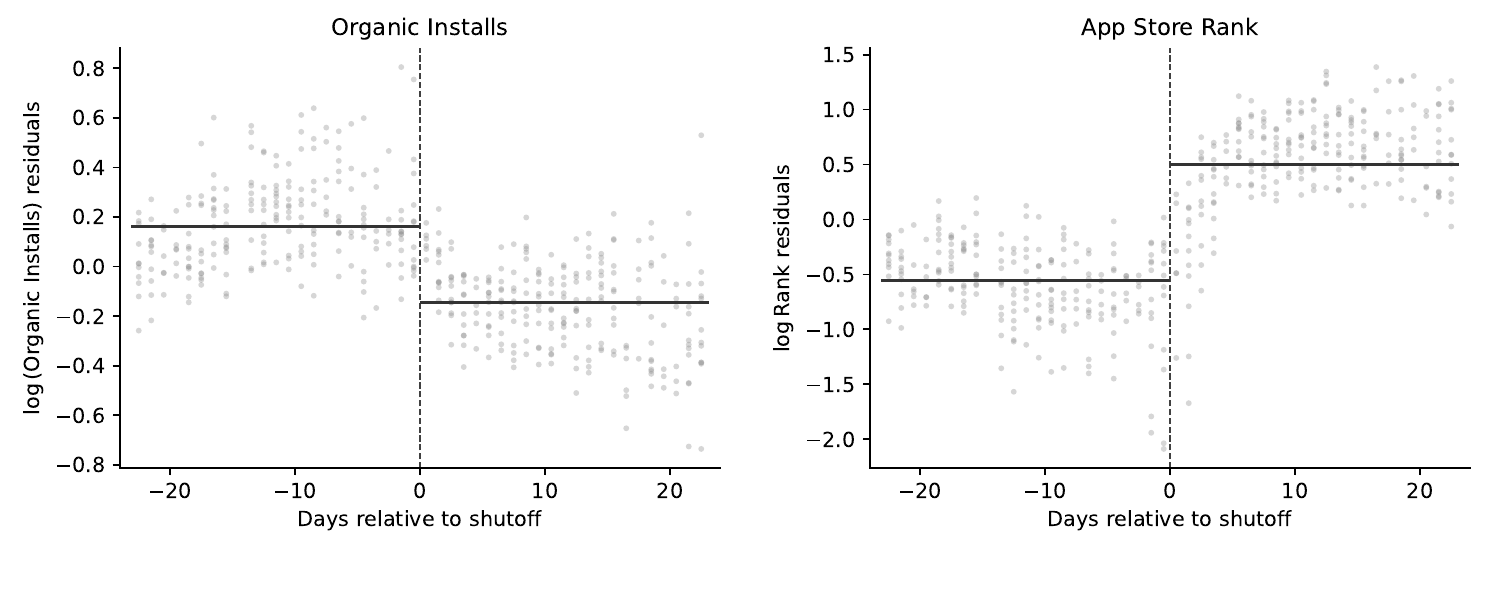}}
\caption{\small Parallel Event Study Discontinuities at Ad Shutoff. \textmd{Residuals after controlling for app-OS and day-of-week fixed effects for $\log$(organic installs) (left) and $\log\text{Rank}$ (right) over an 11-day window around the global shutoff. Each line shows the local constant fit at an 11-day bandwidth. Points are jittered for visualization.}} \label{fig_rdit_dual}
\end{figure}

To evaluate this proposed mechanism ($\text{Ad Spend} \rightarrow \text{App Store Rank} \rightarrow \text{Organic Installs}$), we perform three empirical tests using the daily ranking data from Appfigures for each app's primary category-level Free chart, following the specifications detailed in the previous section. Rank analyses use the subsample with non-missing category rankings, dropping off-chart days that represent roughly 7\% of the post-featuring panel. Figure~\ref{fig_rdit_dual} visualizes the parallel discontinuities in organic installs and store rank at the global shutoff (the corresponding raw, scaled category rank series around the shutoff window is shown in Figure~\ref{fig_rank_zoomed} in the Data section).

First, we replicate our event study design using $\log\text{Rank}_{ijt}$ as the dependent variable, as specified in Equations~\ref{eq_rank_rdit_constant} and \ref{eq_rank_rdit_linear}. Table~\ref{table_rank_rdit} reports local constant event study estimates at representative bandwidths (7, 11, and 23 days). Web Appendix~A reports the full bandwidth grid. We observe a sharp, statistically significant worsening of chart position immediately following the global ad shutoff, mirroring the organic install drop in Figure~\ref{fig_rdit_dual}. At a 23-day bandwidth, $\delta \approx 1.18$ ($p = 0.002$). Interpreting this estimate, an increase of $1.18$ in $\log\text{Rank}_{ijt}$ corresponds to roughly a 3.3-fold deterioration in raw chart position ($\exp(1.18) \approx 3.25$). This confirms causally that the presence of paid ad spend was propping up the games' app store rankings.

\begin{table}[H]
\centering
\begin{tableresize}
\begin{threeparttable}
\begin{tabular}{@{\extracolsep{5pt}}rc}
\\[-1.8ex]\hline
\hline \\[-1.8ex]
& \multicolumn{1}{c}{\textit{Dependent variable:}} \\
\cline{2-2}
\\[-1.8ex] Bandwidth & App Store Rank ($\log\text{Rank}$) \\
\hline \\[-1.8ex]
7 days       & $0.840$\\
            & (0.155; p = .003) \\ \\
11 days       & $1.034$\\
            & (0.176; p = .002) \\ \\
23 days       & $1.183$\\
            & (0.211; p = .002) \\ \\
\hline \\[-1.8ex]
Obs               & 466 \\
\hline
\hline
\end{tabular}
\caption{Event Study Estimates of Ad Shutoff on App Store Rankings} \label{table_rank_rdit}
\end{threeparttable}
\end{tableresize}
\end{table}

Second, we investigate whether yesterday's advertising spend predicts today's store ranking, reflecting the rolling-window nature of the app store's ranking algorithms. Estimating the lagged-effects specification in Equation~\ref{eq_rank_temporal}, we find a negative and statistically significant relationship: each additional \$1,000 of yesterday's spend is associated with a $0.046$-unit decrease in $\log\text{Rank}_{ijt}$ ($p = 0.02$), i.e., a better chart position, roughly two positions at rank~50. This pattern is consistent with category charts weighting recent install velocity in a rolling window. Specifically, yesterday's ad-driven installs keep today's rankings elevated, confirming that the ranking impact of advertising persists beyond the day of the expenditure. Web Appendix~B reports cumulative two- and three-day lag specifications with the same qualitative pattern. The results of this panel analysis are presented in Panel A of Table~\ref{table_rank_mediation}.

Third, we estimate parallel fixed-effects specifications containing both ad spend and app store category rank as predictors, following Equation~\ref{eq_rank_mediation}. This mediation model evaluates whether the contemporaneous association between ad spend and organic installs operates primarily through chart position. If the ranking mechanism is the dominant driver of organic spillovers, then controlling for app store rank should attenuate and absorb the coefficient on ad spend. The results of these joint panel specifications are presented in Panel B of Table~\ref{table_rank_mediation}.

\begin{table}[H]
\centering
\begin{tableresize}
\begin{threeparttable}
\begin{tabular}{@{\extracolsep{5pt}}lcc}
\\[-1.8ex]\hline
\hline \\[-1.8ex]
\multicolumn{3}{l}{\textbf{Panel A: Lagged Effects of Spend on Store Rankings}} \\
\hline \\[-1.8ex]
& \multicolumn{2}{c}{\textit{Dependent variable: $\log\text{Rank}_{ijt}$}} \\
\cline{2-3}
\hline \\[-1.8ex]
$\text{spend}_{ij(t-1)}$ & \multicolumn{2}{c}{$-0.046$} \\
 & \multicolumn{2}{c}{(0.014; p = .02)} \\ \\
\hline \\[-1.8ex]
R$^2$            & \multicolumn{2}{c}{0.847} \\
Obs               & \multicolumn{2}{c}{5475}           \\
\hline \\[-1.8ex]
\multicolumn{3}{l}{\textbf{Panel B: Joint Spend and Store Ranking Specifications (Mediation)}} \\
\hline \\[-1.8ex]
& \multicolumn{2}{c}{\textit{Dependent variable:}} \\
\cline{2-3}
\\[-1.8ex] & Organic Installs & Paid Installs \\
\hline \\[-1.8ex]
$\text{spend}_{ijt}$ & $-1.967$ & $24.308$\\
 & (3.431; p = .59) & (3.234; p = .001) \\ \\
$\log\text{Rank}_{ijt}$ & $-885.795$ & $-1621.668$\\
 & (388.077; p = .07) & (346.947; p = .005) \\ \\
\hline \\[-1.8ex]
R$^2$            & 0.952 & 0.927 \\
Obs               & 5483           & 5483           \\
\hline
\hline
\end{tabular}

\caption{Lagged and Mediation Fixed-Effects Panel Specifications} \label{table_rank_mediation}
\begin{tablenotes}[para,flushleft]
\item \small \textit{Notes:} Rank coefficients in Panel A are changes in $\log\text{Rank}_{ijt}$ per \$1,000 of ad spend. Install coefficients in Panel B are per \$100 of ad spend.
\end{tablenotes}
\end{threeparttable}
\end{tableresize}
\end{table}

For organic installs, we find that adding log rank to the regression almost completely absorbs the spend coefficient. On the same rank subsample, a spend-only specification yields 2.1 organic installs per \$100. In Table~\ref{table_rank_mediation} (Panel B), the spend coefficient falls to $-2.0$ installs per \$100 and becomes statistically insignificant ($p = 0.59$) once log rank is included. At the same time, log rank itself has a strong and marginally significant negative association with organic installs ($\hat{\beta}_2 = -885.80, p = 0.07$). Because standard app store charts are numerically inverted (where lower numbers represent better positions), this negative coefficient confirms that improved chart positions are strongly associated with higher organic install volumes.

For paid installs, a distinct and supportive pattern emerges. Ad spend remains a highly significant and major driver of paid installs, even after controlling for chart position ($\hat{\beta}_1 = 24.31$ installs per \$100, $p = 0.001$). This modest attenuation from the spend-only estimate of 31.7 paid installs per \$100 on the same subsample is consistent with paid conversions being driven primarily by direct, contemporaneous ad delivery and last-touch attribution rather than organic chart discovery. Interestingly, log rank also exhibits a strong negative association with paid installs ($\hat{\beta}_2 = -1621.67, p = 0.005$), suggesting that highly ranked apps may experience higher organic-to-paid conversion rates or enjoy positive synergies on paid acquisition channels. Together, these parallel results are consistent with the mediation hypothesis. Paid ads generate contemporaneous paid installs, but their influence on organic installs is mediated by the app store's algorithmic ranking charts.

\section*{Discussion} \label{section_discussion}

Prior shutoff experiments in search advertising find substitution between paid and organic traffic, but the magnitude varies widely across studies. In app stores, where discovery runs through ranking charts, it remains an open question whether paid install advertising cannibalizes or complements organic installs. We studied an ad spend shutoff experiment conducted by a major mobile game developer. Using a variety of empirical methods, we find no evidence of cannibalization. Rather, our results, consistent across all of our different empirical strategies, indicate that advertising boosts organic installs.

We demonstrate that a ranking mechanism drives this organic lift. Paid install velocity improves store rankings, which in turn drives organic downloads. Prior work shows that chart rankings affect app demand \citep{carare2012bestseller}, that advertising can amplify sales through published ranks \citep{dover2015rankamplification}, and that freemium launches can spill over through enhanced discovery \citep{deng2023freemium}. We extend this literature by identifying paid install advertising as a driver of category rank using shutoff variation. Some organic lift may also stem from unattributed installs (\textit{e.g.}, a consumer sees an ad on one device and installs ``organically'' on another) or word-of-mouth marketing stimulated by ad-driven social proof \citep{rosenfelder2019organiclift}. However, our parallel event studies, lagged ranking analyses, and associational mediation checks are consistent with the ranking mechanism being a primary driver of organic spillovers.

This structural difference in discovery explains why our results differ from prior search ad studies. Previous experiments focused on navigational search queries for well-known brands like eBay and Edmunds, where consumers use search to find a specific website \citep{blake2015consumer, Coviello2017large}. While prior literature has shown that substitution effects are smaller for less popular brands \citep{simonov2018competition}, our results reveal that in mobile ecosystems, the effect actually becomes complementary. This reversal occurs because discovery is heavily mediated by central platform lists rather than navigational search queries. Instead of cannibalizing existing intent, paid advertising drives the app up public ranking charts, boosting organic visibility and subsequent installs. Furthermore, because category rankings represent a public, shared storefront state \citep{dover2015rankamplification, liang2019}, platform incrementality tools such as Meta Lift tests---user-level randomized holdouts that compare ad-eligible and no-ad control users \citep{burtch2025divergent}---may still miss ranking-mediated spillovers. Paid installs from treated users can prop up category charts and boost organic discovery for untreated users as well, violating the stable-unit treatment-value assumption in shared discovery environments. Aggregate, market-level interventions, such as the global shutoff analyzed in this study, are therefore particularly useful for identifying these effects. Related work on multicell experimental designs shows that standard single-cell holdout tests may not recover the estimands needed for intensive-margin decisions such as optimal spend or reach \citep{waisman2025multicell}; our results suggest a complementary limitation when advertising affects a shared market state such as category rankings.

Other secondary factors across digital channels and media types also play a role. Search ads and mobile app install ads are not mutually exclusive, since mobile app install ads include the entire spectrum of digital channels, including search, social media, display, video, and in-app. Exploratory platform-decomposed regressions in Web Appendix~C are consistent with last-touch attribution understating non-search channels relative to search \citep{li2014attributing}, though cross-platform click spillovers are imprecise \citep{golden2017effects, simonov2018competition}.

Another limitation is that, while GameSpace did not strategically choose a date to experiment, confounders during the shutoff window could affect event study estimates. GameSpace also shifted spend forward before the global shutoff, so our estimates compare elevated pre-shutoff spend to near-zero spend rather than typical spend levels. We address this partly by varying bandwidth and reporting local linear estimates in Web Appendix~A, but residual confounding cannot be ruled out. A further limitation is the aggregate nature of our analysis: the shutoff applies across all campaigns and publishers, so we cannot identify campaign- or publisher-level heterogeneity in the experimental effect. Additionally, our store ranking analysis relies on historical ranking archives from Appfigures. While Appfigures is a leading market intelligence provider, these vendor-reported metrics are subject to proprietary collection and smoothing methodologies rather than reflecting first-party app store feeds. There are also category differences across stores (iOS Puzzle Free vs. Android Games Free) necessitated by differing category hierarchies, which may introduce unobserved differences in ranking algorithms and volatility. Lastly, results may not generalize beyond large mobile game developers. Poisson robustness checks in the Web Appendix yield qualitatively similar event study estimates.

Additional limitations stem from the complex nature of digital advertising. As noted by \cite{bagwell2007economic}, advertising decisions are generally endogenous, which can bias fixed effects panel estimates. However, several factors for our context may suggest that our panel results are informative: day-to-day budget allocations were often arbitrary at the app level, and app-OS fixed effects absorb deliberate cross-app differences. The context of mobile apps introduces further complexity, including location \citep{ghose2013mobile, luo2014mobile, fang2015contemporaneous, fong2015geo}, product utility \citep{bart2014products}, and personalization \citep{zhang2019personalized}. Consumer heterogeneity also has substantial implications, especially in the mobile gaming industry where consumer spending is highly skewed with a small minority generating a majority of revenue.

Our study provides empirical evidence of positive organic spillovers in mobile app install advertising under an ad spend shutoff. Paid advertising is more effective than attributed paid install numbers suggest, so developers who optimize on paid installs alone may be underinvesting in marketing.

\section*{Acknowledgments}
The authors gratefully acknowledge John Horton, Catherine Tucker, Dean Eckles, and Sunil Wattal for their comments on the paper and the support of the MIT Initiative on the Digital Economy.

\clearpage
\singlespacing
\bibliographystyle{plainnat}
\bibliography{references}

\clearpage
\doublespacing
\appendix
\renewcommand{\thetable}{A\arabic{table}}
\renewcommand{\thefigure}{A\arabic{figure}}
\setcounter{table}{0}
\setcounter{figure}{0}

\section*{Web Appendix A: Robustness Checks} \label{section_web_appendix_a}

\begin{figure}[H]
\centering
{\includegraphics[width=\linewidth]{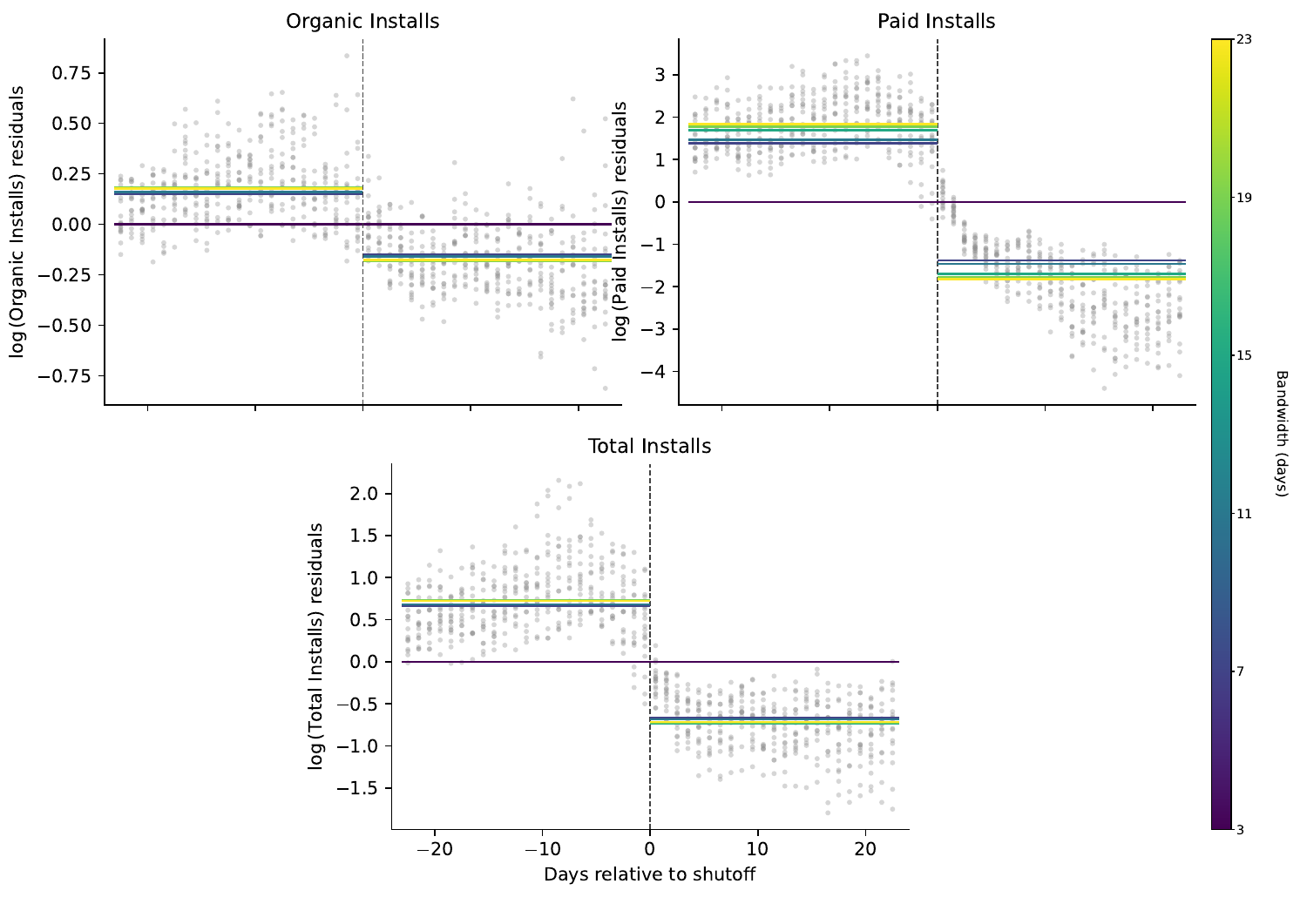}}
\caption{Local Constant Event Study Fits across Bandwidths. \textmd{Same design as the main-text install event study figure, with local constant fits at bandwidths from 3 to 23 days (color scale).}} \label{fig_es_bandwidths}
\end{figure}

\begin{figure}[H]
\centering
{\includegraphics[width=\linewidth]{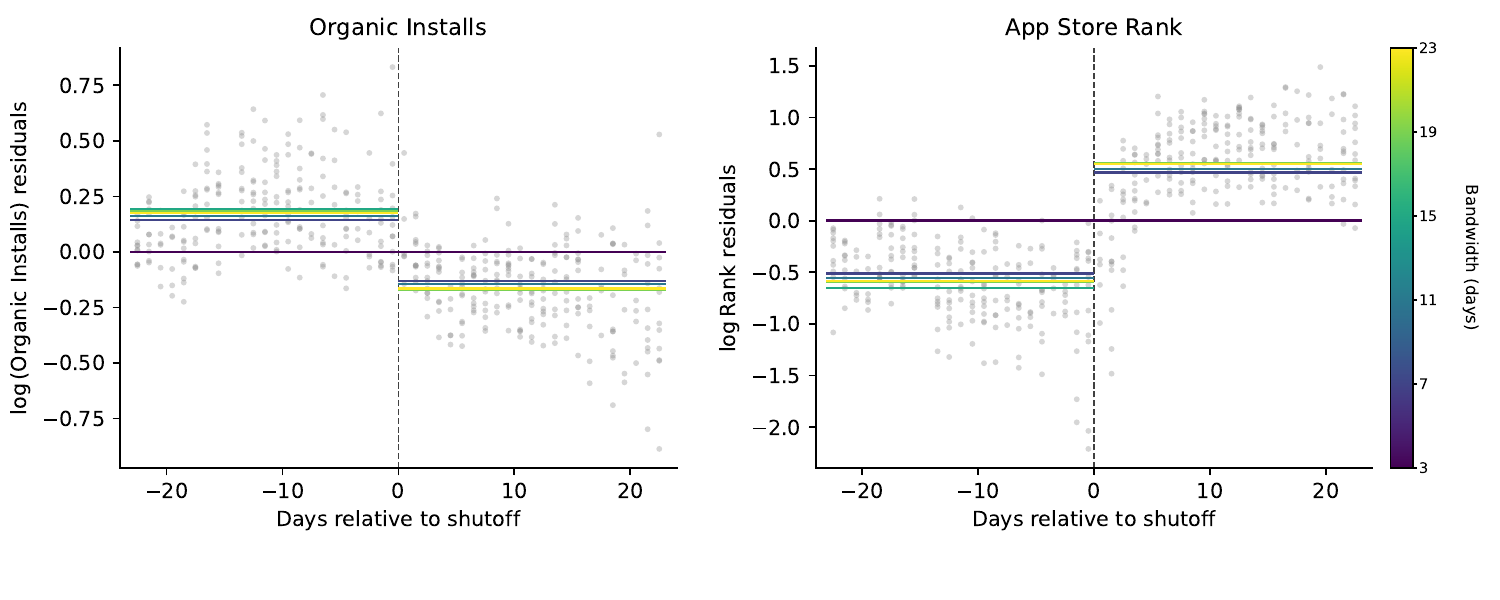}}
\caption{Local Constant Parallel Event Study Discontinuities across Bandwidths. \textmd{Same design as the main-text parallel event study figure, with local constant fits at bandwidths from 3 to 23 days (color scale).}} \label{fig_rdit_dual_bandwidths}
\end{figure}

\begin{figure}[H]
\centering
{\includegraphics[width=0.9\linewidth]{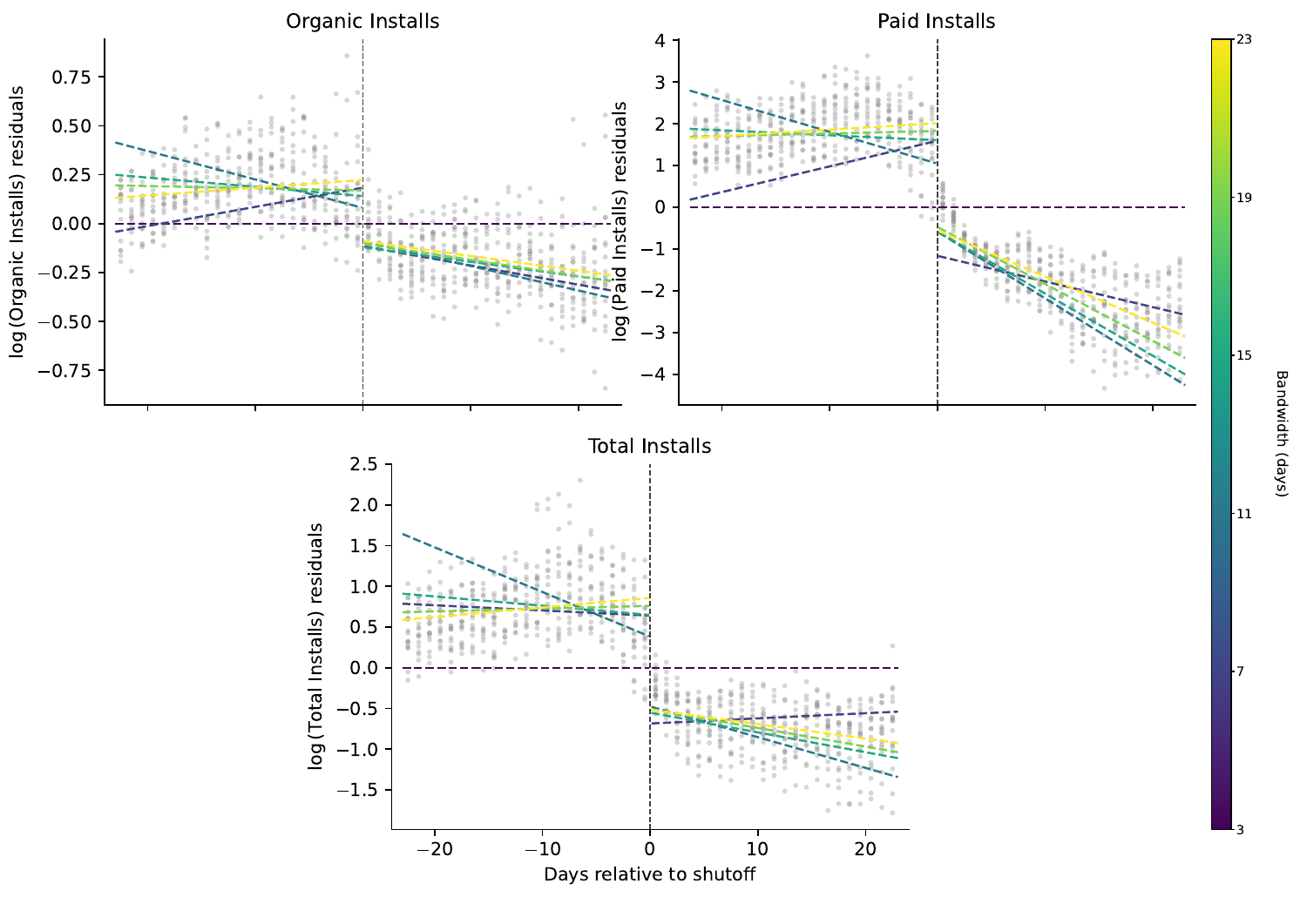}}
\caption{Local Linear Event Study Fits at Ad Shutoff. \textmd{Same design as the main-text install event study figure, with local linear fits at bandwidths from 3 to 23 days (color scale).}} \label{fig_es_linear}
\end{figure}

Our main text reports local constant event study estimates at representative bandwidths (7, 11, and 23 days). Figures~\ref{fig_es_bandwidths} and~\ref{fig_rdit_dual_bandwidths} report the corresponding local constant fits across bandwidths from 3 to 23 days. Tables~\ref{table_es_all_bandwidths} and~\ref{table_rank_rdit_all_bandwidths} report the full local constant grids. Table~\ref{table_es_linear} and Figure~\ref{fig_es_linear} report the corresponding local linear estimates. Under the local linear specification, organic-install discontinuities grow increasingly negative as bandwidth increases, but the qualitative conclusion---a sharp drop at shutoff---is unchanged. The local linear fits are also useful diagnostics: pre-shutoff slopes turn from negative to positive as bandwidth widens, consistent with GameSpace shifting spend forward before the shutoff, and post-shutoff slopes remain negative across bandwidths, consistent with advertising persistence.

\begin{table}[H]
\centering
\begin{tableresize}
\begin{threeparttable}
\begin{tabular}{@{\extracolsep{5pt}}rccc}
\\[-1.8ex]\hline
\hline \\[-1.8ex]
& \multicolumn{3}{c}{\textit{Dependent variable:}} \\
\cline{2-4}
\\[-1.8ex] Bandwidth & \multicolumn{1}{c}{Organic Installs} & \multicolumn{1}{c}{Paid Installs} & \multicolumn{1}{c}{Total Installs}  \\
\hline \\[-1.8ex]
3 days       & $-0.227$ & $-2.229$ & $-1.155$\\
            & (0.061; p = .01) & (0.162; p = .001) & (0.158; p = .001) \\ \\
7 days       & $-0.305$ & $-2.732$ & $-1.311$\\
            & (0.061; p = .004) & (0.203; p = .001) & (0.179; p = .001) \\ \\
11 days       & $-0.319$ & $-3.013$ & $-1.368$\\
            & (0.066; p = .005) & (0.235; p = .001) & (0.189; p = .001) \\ \\
15 days       & $-0.329$ & $-3.316$ & $-1.400$\\
            & (0.070; p = .005) & (0.275; p = .001) & (0.197; p = .001) \\ \\
19 days       & $-0.347$ & $-3.551$ & $-1.440$\\
            & (0.075; p = .006) & (0.301; p = .001) & (0.204; p = .001) \\ \\
23 days       & $-0.359$ & $-3.709$ & $-1.460$\\
            & (0.081; p = .007) & (0.311; p = .001) & (0.208; p = .001) \\ \\
\hline \\[-1.8ex]
Obs               & 825 & 825 & 825 \\
\hline
\hline
\end{tabular}

\caption{Local Constant Event Study Estimates of Ad Shutoff Effects (All Bandwidths)}  \label{table_es_all_bandwidths}
\end{threeparttable}
\end{tableresize}
\end{table}

\begin{table}[H]
\centering
\begin{tableresize}
\begin{threeparttable}
\begin{tabular}{@{\extracolsep{5pt}}rc}
\\[-1.8ex]\hline
\hline \\[-1.8ex]
& \multicolumn{1}{c}{\textit{Dependent variable:}} \\
\cline{2-2}
\\[-1.8ex] Bandwidth & App Store Rank ($\log\text{Rank}$) \\
\hline \\[-1.8ex]
3 days       & $0.449$\\
            & (0.215; p = .09) \\ \\
7 days       & $0.840$\\
            & (0.155; p = .003) \\ \\
11 days       & $1.034$\\
            & (0.176; p = .002) \\ \\
15 days       & $1.117$\\
            & (0.190; p = .002) \\ \\
19 days       & $1.156$\\
            & (0.199; p = .002) \\ \\
23 days       & $1.183$\\
            & (0.211; p = .002) \\ \\
\hline \\[-1.8ex]
Obs               & 466 \\
\hline
\hline
\end{tabular}
\caption{Local Constant Event Study Estimates of Ad Shutoff on App Store Rankings (All Bandwidths)} \label{table_rank_rdit_all_bandwidths}
\end{threeparttable}
\end{tableresize}
\end{table}

\begin{table}[H]
\centering
\begin{tableresize}
\begin{threeparttable}
\begin{tabular}{@{\extracolsep{5pt}}rccc}
\\[-1.8ex]\hline
\hline \\[-1.8ex]
& \multicolumn{3}{c}{\textit{Dependent variable:}} \\
\cline{2-4}
\\[-1.8ex] Bandwidth & \multicolumn{1}{c}{Organic Installs} & \multicolumn{1}{c}{Paid Installs} & \multicolumn{1}{c}{Total Installs}  \\
\hline \\[-1.8ex]
3 days       & $-0.189$ & $-1.910$ & $-1.124$\\
            & (0.077; p = .06) & (0.213; p = .001) & (0.187; p = .002) \\ \\
7 days       & $-0.240$ & $-2.142$ & $-1.255$\\
            & (0.075; p = .02) & (0.154; p = .001) & (0.175; p = .001) \\ \\
11 days       & $-0.303$ & $-2.355$ & $-1.327$\\
            & (0.074; p = .009) & (0.196; p = .001) & (0.193; p = .001) \\ \\
15 days       & $-0.321$ & $-2.361$ & $-1.370$\\
            & (0.075; p = .008) & (0.161; p = .001) & (0.195; p = .001) \\ \\
19 days       & $-0.316$ & $-2.459$ & $-1.380$\\
            & (0.075; p = .008) & (0.170; p = .001) & (0.199; p = .001) \\ \\
23 days       & $-0.322$ & $-2.646$ & $-1.410$\\
            & (0.078; p = .009) & (0.209; p = .001) & (0.208; p = .001) \\ \\
\hline \\[-1.8ex]
Obs               & 825 & 825 & 825 \\
\hline
\hline
\end{tabular}

\caption{Local Linear Event Study Estimates of Ad Shutoff Effects}  \label{table_es_linear}
\end{threeparttable}
\end{tableresize}
\end{table}

\begin{figure}[H]
\centering
{\includegraphics[width=\linewidth]{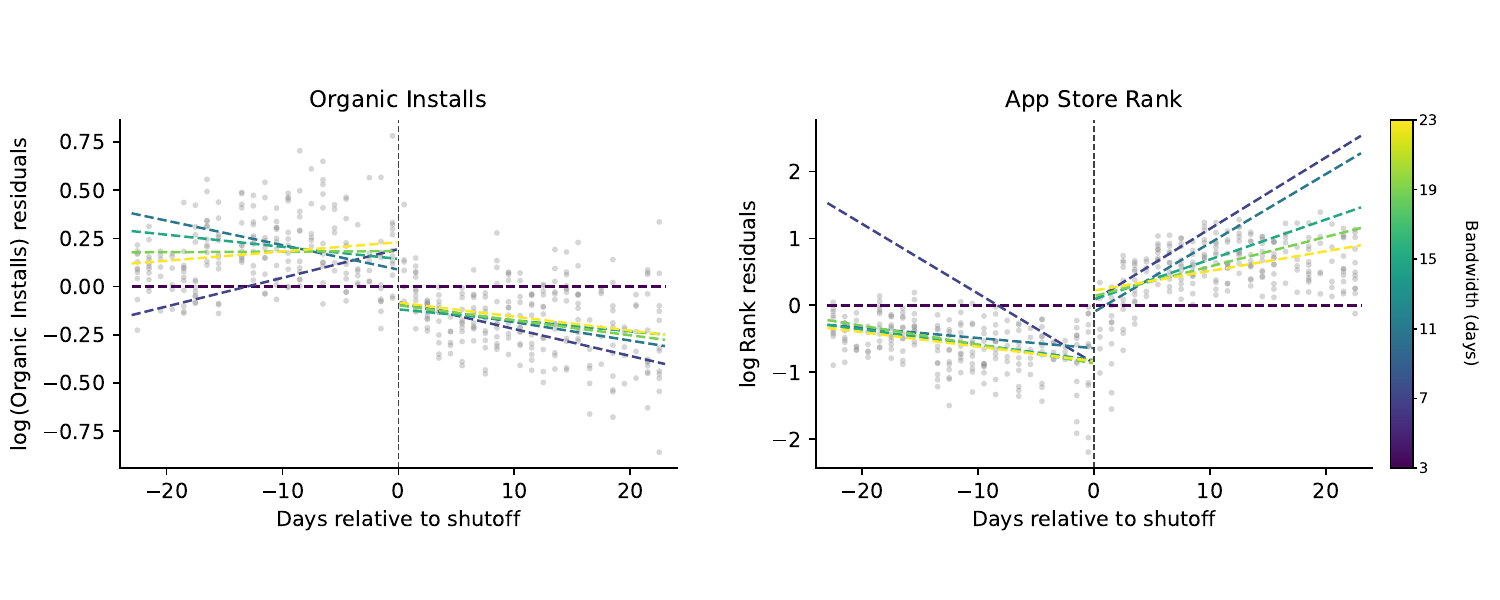}}
\caption{Local Linear Parallel Event Study Discontinuities. \textmd{Same design as the main-text parallel event study figure, with local linear fits at bandwidths from 3 to 23 days (color scale).}} \label{fig_rdit_dual_linear}
\end{figure}

Table~\ref{table_rank_rdit_linear} reports local linear event study estimates for app store rankings. Estimates are stable across bandwidths of 7 days and above.

\begin{table}[H]
\centering
\begin{tableresize}
\begin{threeparttable}
\begin{tabular}{@{\extracolsep{5pt}}rc}
\\[-1.8ex]\hline
\hline \\[-1.8ex]
& \multicolumn{1}{c}{\textit{Dependent variable:}} \\
\cline{2-2}
\\[-1.8ex] Bandwidth & App Store Rank ($\log\text{Rank}$) \\
\hline \\[-1.8ex]
3 days       & $0.358$\\
            & (0.249; p = .21) \\ \\
7 days       & $0.417$\\
            & (0.244; p = .15) \\ \\
11 days       & $0.688$\\
            & (0.154; p = .007) \\ \\
15 days       & $0.858$\\
            & (0.139; p = .002) \\ \\
19 days       & $0.976$\\
            & (0.142; p = .001) \\ \\
23 days       & $1.067$\\
            & (0.143; p = .001) \\ \\
\hline \\[-1.8ex]
Obs               & 466 \\
\hline
\hline
\end{tabular}
\caption{Local Linear Event Study Estimates of Ad Shutoff on App Store Rankings} \label{table_rank_rdit_linear}
\end{threeparttable}
\end{tableresize}
\end{table}

Install counts are nonnegative integers. To ensure robustness of our event study estimates, we re-estimated the local constant and local linear specifications using Poisson regression as an alternative to the main log-linear OLS specification. Table~\ref{table_es_poisson} reports Poisson discontinuity estimates. Results are qualitatively similar to the main event study table. At a bandwidth of 11 days, the local constant Poisson discontinuity for organic installs is approximately $-0.29$, compared with $-0.32$ in the main specification.

\begin{table}[H]
\centering
\begin{tableresize}
\begin{threeparttable}
\begin{tabular}{@{\extracolsep{5pt}}rcccccc}
\\[-1.8ex]\hline
\hline \\[-1.8ex]
& \multicolumn{6}{c}{\textit{Dependent variable:}} \\
\cline{2-7}
\\[-1.8ex] & \multicolumn{2}{c}{Organic Installs} & \multicolumn{2}{c}{Paid Installs} & \multicolumn{2}{c}{Total Installs}  \\
\\[-1.8ex] Bandwidth & L. Constant & L. Linear & L. Constant & L. Linear & L. Constant & L. Linear \\
\hline \\[-1.8ex]
3 days   & $-0.219$ & $-0.166$ & $-2.029$ & $-1.787$ & $-1.047$ & $-1.034$\\
            & (0.051; p = .001) & (0.080; p = .04) & (0.122; p = .001) & (0.247; p = .001) & (0.131; p = .001) & (0.196; p = .001) \\ \\
7 days   & $-0.274$ & $-0.209$ & $-2.406$ & $-1.949$ & $-1.180$ & $-1.122$\\
            & (0.060; p = .001) & (0.075; p = .005) & (0.166; p = .001) & (0.200; p = .001) & (0.157; p = .001) & (0.181; p = .001) \\ \\
11 days   & $-0.285$ & $-0.250$ & $-2.577$ & $-2.095$ & $-1.222$ & $-1.183$\\
            & (0.064; p = .001) & (0.076; p = .001) & (0.182; p = .001) & (0.212; p = .001) & (0.163; p = .001) & (0.195; p = .001) \\ \\
15 days   & $-0.292$ & $-0.266$ & $-2.748$ & $-2.150$ & $-1.255$ & $-1.217$\\
            & (0.068; p = .001) & (0.079; p = .001) & (0.203; p = .001) & (0.205; p = .001) & (0.170; p = .001) & (0.197; p = .001) \\ \\
19 days   & $-0.299$ & $-0.270$ & $-2.871$ & $-2.215$ & $-1.281$ & $-1.236$\\
            & (0.070; p = .001) & (0.082; p = .001) & (0.215; p = .001) & (0.208; p = .001) & (0.175; p = .001) & (0.201; p = .001) \\ \\
23 days   & $-0.309$ & $-0.268$ & $-2.973$ & $-2.281$ & $-1.300$ & $-1.257$\\
            & (0.072; p = .001) & (0.084; p = .001) & (0.224; p = .001) & (0.214; p = .001) & (0.177; p = .001) & (0.206; p = .001) \\ \\
\hline \\[-1.8ex]
Obs               & 825 & 825 & 825 & 825 & 825 & 825 \\
\hline
\hline
\end{tabular}

\caption{Poisson Robustness: Event Study Estimates}  \label{table_es_poisson}
\end{threeparttable}
\end{tableresize}
\end{table}

We also estimated fixed-effects specifications with lagged ad spend to investigate carryover effects beyond the contemporaneous model in the main text. Table~\ref{table_panel_lags} reports contemporaneous and one-day-lagged spend for organic, paid, and total installs. Two-day lags were insignificant and are omitted. Lagged spend coefficients for organic installs are not statistically significant. For paid installs, every \$100 spent is associated with approximately 29.7 same-day paid installs and 2.8 paid installs on the following day ($p = 0.05$).

\begin{table}[H]
\centering
\begin{tableresize}
\begin{threeparttable}
\begin{tabular}{@{\extracolsep{2pt}}lcccccc}
\\[-1.8ex]\hline
\hline \\[-1.8ex]
& \multicolumn{6}{c}{\textit{Dependent variable:}} \\
\cline{2-7}
\\[-1.8ex] & \multicolumn{2}{c}{Organic Installs} & \multicolumn{2}{c}{Paid Installs} & \multicolumn{2}{c}{Total Installs} \\
\\[-1.8ex] & Contemporaneous & + Lag & Contemporaneous & + Lag & Contemporaneous & + Lag \\
\hline \\[-1.8ex]
$\text{spend}_{ijt}$  & $2.177$ & $2.521$ & $32.295$ & $29.656$ & $34.472$ & $32.176$\\
 & (3.042; p = .51) & (1.209; p = .09) & (3.867; p = .001) & (3.225; p = .001) & (6.638; p = .003) & (4.152; p = .001)\\ \\
$\text{spend}_{ij(t-1)}$  &  & $-0.351$ &  & $2.823$ &  & $2.472$\\
 &  & (1.955; p = .86) &  & (1.092; p = .05) &  & (3.002; p = .45)\\ \\
\hline \\[-1.8ex]
R$^2$            & 0.942 & 0.942 & 0.919 & 0.920 & 0.936 & 0.936 \\
Obs               & 5829           & 5829           & 5829           & 5829           & 5829           & 5829           \\
\hline
\hline
\end{tabular}

\caption{Lagged Fixed-Effects Specifications: Installs per \$100 of Ad Spend} \label{table_panel_lags}
\end{threeparttable}
\end{tableresize}
\end{table}

\clearpage
\section*{Web Appendix B: Rank Mechanism Robustness} \label{section_web_appendix_b}

As described in the main text, yesterday's advertising spend predicts today's $\log\text{Rank}$: each additional \$1,000 of yesterday's spend is associated with a $0.046$-unit decrease in $\log\text{Rank}_{ijt}$ ($p = 0.02$). Table~\ref{table_rank_lags} extends this specification cumulatively to two- and three-day lags, reporting coefficients per \$1,000 of ad spend. Lagged spend coefficients remain negative and statistically significant through three days, consistent with category charts weighting recent install velocity over a rolling window of several days.

\begin{table}[H]
\centering
\begin{tableresize}
\begin{threeparttable}
\begin{tabular}{@{\extracolsep{5pt}}lccc}
\\[-1.8ex]\hline
\hline \\[-1.8ex]
& \multicolumn{3}{c}{\textit{Dependent variable: $\log\text{Rank}_{ijt}$}} \\
\cline{2-4}
\\[-1.8ex] & 1-Day Lag & 2-Day Lags & 3-Day Lags \\
\hline \\[-1.8ex]
$\text{spend}_{ij(t-1)}$ & $-0.046$ & $-0.022$ & $-0.021$\\
 & (0.014; p = .02) & (0.007; p = .03) & (0.007; p = .03) \\ \\
$\text{spend}_{ij(t-2)}$ &  & $-0.026$ & $-0.009$\\
 &  & (0.007; p = .01) & (0.002; p = .007) \\ \\
$\text{spend}_{ij(t-3)}$ &  &  & $-0.020$\\
 &  &  & (0.008; p = .05) \\ \\
\hline \\[-1.8ex]
R$^2$            & 0.851 & 0.858 & 0.862 \\
Obs               & 5451 & 5451 & 5451 \\
\hline
\hline
\end{tabular}

\caption{Lagged Fixed-Effects Specifications: Ad Spend and Store Rankings} \label{table_rank_lags}
\begin{tablenotes}[para,flushleft]
\item \small \textit{Notes:} Coefficients are changes in $\log\text{Rank}_{ijt}$ per \$1,000 of ad spend.
\end{tablenotes}
\end{threeparttable}
\end{tableresize}
\end{table}

To assess sensitivity to the rank functional form, Table~\ref{table_rank_transform_robustness} replicates the core rank-mechanism specifications using alternative monotonic transformations that code higher values as better chart positions: $-\log(\mathrm{Rank})$, $-\mathrm{Rank}$, and $1/\mathrm{Rank}$. Across all three transformations, the event study shutoff discontinuity remains negative and statistically significant, and we find substantial spend attenuation of 35.9\% to 64.9\% when rank is included, confirming that the ranking-mediation pattern is highly robust to alternative functional form assumptions.

Appfigures does not report a rank when an app falls off the tracked category chart. Our main analyses therefore drop those off-chart days (about 7\% of the post-featuring panel). Table~\ref{table_rank_off_chart_robustness} assesses sensitivity by instead coding all off-chart days as rank~401, just below Appfigures' default top-400 cutoff. The event study rank discontinuity and lagged rank effect remain statistically significant with the same signs under both treatments, and we observe robust spend attenuation of 32.3\% under the censored treatment compared to 64.9\% in the observed-only sample, confirming that our findings are not driven by the exclusion of off-chart days.

\begin{table}[H]
\centering
\begin{tableresize}
\begin{threeparttable}
\begin{tabular}{@{\extracolsep{5pt}}lcc}
\\[-1.8ex]\hline
\hline \\[-1.8ex]
& \multicolumn{2}{c}{\textit{Off-chart day treatment}} \\
\cline{2-3}
\\[-1.8ex] & Observed only & Censored at 401 \\
\hline \\[-1.8ex]
Event study $\delta$ (23-day bw) & $1.183$ & $0.878$\\
 & (0.211; p = .002) & (0.174; p = .004) \\ \\
Lagged effects ($\text{spend}_{t-1} \rightarrow \log\text{Rank}_t$, per \$1,000) & $-0.046$ & $-0.039$\\
 & (0.014; p = .02) & (0.011; p = .02) \\ \\
Spend attenuation (\%, adding rank) & $64.9\%$ & $32.3\%$\\
\hline \\[-1.8ex]
Obs (event study / panel) & 466 / 5475 & 549 / 5817\\
\hline
\hline
\end{tabular}

\caption{Robustness to Off-Chart Day Treatment} \label{table_rank_off_chart_robustness}
\end{threeparttable}
\end{tableresize}
\end{table}

\begin{table}[H]
\centering
\begin{tableresize}
\begin{threeparttable}
\begin{tabular}{@{\extracolsep{5pt}}lccc}
\\[-1.8ex]\hline
\hline \\[-1.8ex]
& \multicolumn{3}{c}{\textit{Rank transformation (higher = better)}} \\
\cline{2-4}
\\[-1.8ex] & $-\log(\mathrm{Rank})$ & $-\mathrm{Rank}$ & $1/\mathrm{Rank}$ \\
\hline \\[-1.8ex]
Event study $\delta$ (23-day bw) & $-1.183$ & $-100.414$ & $-0.026$\\
 & (0.211; p = .002) & (22.888; p = .007) & (0.009; p = .03) \\ \\
Spend attenuation (\%, adding rank) & $64.9\%$ & $38.6\%$ & $35.9\%$\\
\hline \\[-1.8ex]
Obs (event study / panel) & 466 / 5475 & 466 / 5475 & 466 / 5475\\
\hline
\hline
\end{tabular}

\caption{Robustness of Rank Mechanism to Alternative Rank Transformations} \label{table_rank_transform_robustness}
\end{threeparttable}
\end{tableresize}
\end{table}

\clearpage
\section*{Web Appendix C: Platform-Decomposed Regressions} \label{section_web_appendix_c}

Since many firms allocate marketing spend across multiple platforms, we report exploratory associations between platform-level spend and installs. Because our shutoff varies aggregate spend rather than at the platform level, we cannot identify causal platform-level effects. We nonetheless decompose the main fixed-effects specification by major ad platforms and examine cross-platform click spillovers.

\noindent\textit{Spend by Platform.}\par
\noindent To examine the impact of ad spend on installs by platform, we superscript spend terms in Equation~\ref{eq_spend_decomp}:
\begin{equation}
Y_{ijt} = \sum_{k \in K} \beta^k \text{spend}_{ijt}^k + \pi_{ijt}  + \epsilon_{ijt}
\label{eq_spend_decomp}
\end{equation}
where $k\in K$ is either Google or Facebook, the two platforms accounting for 48\% of ad spend, and $\pi_{ijt}$ is the app-OS-time fixed effect from the main text.

\noindent\textit{Impressions by Platform.}\par
\noindent Cross-platform spillovers can affect the estimation of ad effectiveness across publishers \citep{li2014attributing}. To examine the impact of impressions on clicks by platform, we estimate the effects of $\text{impressions}_{ijt}^k$ instead of $\text{spend}_{ijt}^k$. To estimate cross-platform effects, we denote an additional superscript $k'\in K$ on the outcome variable to distinguish it from the superscript $k$ on the regressor. Thus, we obtain Equation~\ref{eq_impressions_decomp}:
\begin{equation}
Y_{ijt}^{k'} = \sum_{k \in K} \beta^k \text{impressions}_{ijt}^k + \alpha_{ij} + \tau_{t}  + \epsilon_{ijt} \text{.}
\label{eq_impressions_decomp}
\end{equation}
The outcome $Y_{ijt}^{k'}$ includes clicks on Google, clicks on Facebook, clicks on all other platforms, and total clicks on all platforms. We used app-OS entity effects $\alpha_{ij}$ and time effects $\tau_t$ because the app-OS-day fixed effect $\pi_{ijt}$ leads to unstable estimates.

When we estimated Equation~\ref{eq_spend_decomp}, we found heterogeneous associations across platforms (Table~\ref{table_panel_platform}). Facebook spend is associated with the largest paid and total install response, at 43.5 and 53.0 installs per \$100, respectively, and with 9.5 additional organic installs per \$100 ($p = 0.09$). Relative to attributed paid installs alone, the Facebook total-install coefficient implies roughly 22\% more installs per dollar than last-touch attribution would credit. Google and other paid channels show null organic associations in this joint specification, while paid-install responses remain large for all three channels.

\begin{table}[H]
\centering
\begin{tableresize}
\begin{threeparttable}
\begin{tabular}{@{\extracolsep{5pt}}lccc}
\\[-1.8ex]\hline
\hline \\[-1.8ex]
& \multicolumn{3}{c}{\textit{Dependent variable:}} \\
\cline{2-4}
\\[-1.8ex] & Organic Installs & Paid Installs & Total Installs \\
\hline \\[-1.8ex]
$\text{spend}^{\text{google}}_{ijt}$ & $-0.211$ & $26.757$ & $26.546$\\
 & (2.455; p = .93) & (8.226; p = .02) & (10.645; p = .05) \\ \\
$\text{spend}^{\text{facebook}}_{ijt}$ & $9.519$ & $43.450$ & $52.969$\\
 & (4.485; p = .09) & (9.478; p = .006) & (13.882; p = .01) \\ \\
$\text{spend}^{\text{others}}_{ijt}$ & $-1.757$ & $27.655$ & $25.898$\\
 & (2.953; p = .58) & (3.060; p = .001) & (4.169; p = .002) \\ \\
\hline \\[-1.8ex]
Obs               & 5826 & 5826 & 5826 \\
\hline
\hline
\end{tabular}

\caption{Platform-Decomposed Install Effects per \$100 of Ad Spend} \label{table_panel_platform}
\end{threeparttable}
\end{tableresize}
\end{table}

Moreover, ad effectiveness on individual platforms can be impacted by cross-platform spillovers in impressions and the selected attribution model. To measure cross-platform spillover, we estimated Equation~\ref{eq_impressions_decomp} where the outcome is the number of clicks on platform $k'$ and the coefficients are per 1,000 impressions. On-platform effects are most salient along the diagonal of Table~\ref{table_panel_cross_platform_clicks}. Google impressions generate about 11.9 Google clicks per 1,000 impressions, while Facebook impressions generate about 8.8 Facebook clicks per 1,000 impressions.

\begin{table}[H]
\centering
\begin{tableresize}
\begin{threeparttable}
\begin{tabular}{@{\extracolsep{5pt}}lcccc}
\\[-1.8ex]\hline
\hline \\[-1.8ex]
& \multicolumn{4}{c}{\textit{Dependent variable:}} \\
\cline{2-5}
\\[-1.8ex] & Google Clicks & Facebook Clicks & Other Clicks & Total Clicks \\
\hline \\[-1.8ex]
$\text{impressions}^{\text{google}}_{ijt}$ & $11.914$ & $-0.358$ & $4.611$ & $16.166$\\
 & (2.744; p = .007) & (0.246; p = .21) & (2.954; p = .18) & (4.581; p = .02) \\ \\
$\text{impressions}^{\text{facebook}}_{ijt}$ & $1.507$ & $8.755$ & $0.561$ & $10.824$\\
 & (1.218; p = .27) & (0.632; p = .001) & (5.154; p = .92) & (4.504; p = .06) \\ \\
$\text{impressions}^{\text{others}}_{ijt}$ & $-0.985$ & $-0.451$ & $25.857$ & $24.421$\\
 & (0.313; p = .03) & (0.192; p = .07) & (9.768; p = .05) & (9.547; p = .05) \\ \\
\hline \\[-1.8ex]
Obs               & 5829 & 5829 & 5829 & 5829 \\
\hline
\hline
\end{tabular}

\caption{Cross-Platform Spillover from Impressions to Clicks} \label{table_panel_cross_platform_clicks}
\begin{tablenotes}[para,flushleft]
\item \small \textit{Notes:} Coefficients are clicks per 1,000 impressions.
\end{tablenotes}
\end{threeparttable}
\end{tableresize}
\end{table}

For cross-platform effects, Facebook impressions are associated with 10.8 total clicks per 1,000 impressions ($p = 0.06$), but spillovers to Google or other channels are small and statistically insignificant. Google impressions are associated with 16.2 total clicks per 1,000 impressions ($p = 0.02$). One possible explanation for off-platform click patterns is that some consumers who see an app on Facebook later search for it on Google, as suggested by \cite{li2014attributing}, though we do not find a precise Facebook-to-Google click spillover in this specification. Regardless of the mechanism, the results suggest that cross-platform dynamics may further affect attribution-based estimates of ad effectiveness on some platforms. This issue is especially salient with last-touch attribution, used by GameSpace, which would credit the install only to the platform where the user ultimately clicked, rather than the platforms that delivered earlier impressions. Alternative attribution models, such as time decay or data-driven models, may help reduce such inaccurate estimation of ad effectiveness on some platforms (Google Ads offers time-decay and data-driven alternatives).

\end{document}